\documentclass[12pt]{article}

\usepackage[letterpaper,margin=1in]{geometry}
\usepackage{amsmath}
\usepackage{amssymb}
\usepackage[version=3]{mhchem}
\usepackage{graphicx}
\usepackage[numbers,sort&compress]{natbib}
\usepackage{xurl}
\usepackage[colorlinks=true,linkcolor=blue,citecolor=blue,urlcolor=blue]{hyperref}
\title{Interfacial Magnetotransport in a \ce{NiI2}/Graphene Heterostructure}
\author{Stasiu T. Chyczewski, Xiaotong Xu, and Wenjuan Zhu\thanks{Corresponding author. wjzhu@illinois.edu}\\
\small Department of Electrical and Computer Engineering, University of Illinois Urbana-Champaign\\
\small 208 N Wright St, Urbana, Illinois 61801, USA\\
\small stasiuc2@illinois.edu; xx28@illinois.edu; wjzhu@illinois.edu}
\date{}

\begin{document}

\maketitle

\begin{abstract}
We investigate magnetotransport in a van der Waals heterostructure composed of monolayer graphene and the insulating helical antiferromagnet \ce{NiI2}. While \ce{NiI2} is highly resistive and thus poorly suited for direct transport measurements, we demonstrate that magnetotransport in an adjacent graphene layer provides an electrical readout of magnetic-state-dependent interfacial behavior. Most notably, first-harmonic longitudinal magnetoresistance under in-plane magnetic fields exhibits large, anisotropic low-field peaks that are absent from a monolayer graphene/h-BN control device and are suppressed above the multiferroic transition temperature of \ce{NiI2}. Temperature-dependent harmonic measurements provide complementary evidence: the second-harmonic resistance shows the clearest nonlinear contrast relative to the control device, while the third harmonic contains a larger generic nonlinear and thermal background that is nevertheless modified in the heterostructure. These results demonstrate that graphene-based transport measurements offer a sensitive, non-invasive probe of magnetic phase behavior in electrically insulating van der Waals magnets, opening routes toward spintronic devices based on insulating vdW multiferroics.
\end{abstract}

\section{Introduction}\label{sec1}

Spintronics based on antiferromagnets (AFMs) has recently been recognized as promising for next-generation device applications. Antiferromagnet-based devices hold several advantages over their ferromagnetic counterparts, including resistance to external magnetic perturbations, the absence of stray fields (allowing for denser memory and better security), and the ability to break into the terahertz frequency regime~\cite{RevModPhys.90.015005,RN596,mnbased_review}. While collinear AFMs were first studied for their usage as a source of exchange bias in magnetic heterostructures~\cite{RN88}, they have since been explored as a storage medium in their own right. For example, bits made of simple planar devices using metallic, collinear AFMs have been demonstrated at room temperature and at terahertz speeds.~\cite{RN169,RN597,RN598}. While these findings have advanced collinear AFMs as viable spintronic materials, attention has also begun to shift toward non-collinear systems.

Unlike collinear AFM whose spins alternate along a single axis, non-collinear AFMs have spin arrangements at different angles which still cancel (or leave a very small) net magnetization~\cite{RN596,mnbased_review}. Examples of non-collinear magnet types include triangular, canted, and helical spin arrangements. Non-collinear AFMs can have several advantages, such as large anomalous Hall effects (AHE) and multiferroic behavior~\cite{RN599,RN600}. Beyond conducting AFMs, non-collinear insulating AFMs are of great interest for multiferroic devices. Magnetoelectric coupling in AFMs has long been known, with such behavior reported in the canted AFM \ce{Cr2O3} over half a century ago, and newer work continuing to exploit it for device applications~\cite{first_cr2o3,RN601}. More recently, the multiferroic properties of helical AFMs have been explored. Demonstrations include electrical control of spin helicity and magnon transport for device applications~\cite{helicity_control_2007,RN602}. The range of electrical control mechanisms and phenomena in non-collinear AFMs positions them as attractive candidates for next-generation spintronic and multiferroic devices.

In addition to the already rich library of 3D antiferromagnetic materials, the emergence of van der Waals (vdW) magnets (both ferromagnets and antiferromagnets) brings new possibilities to exploit low dimensionality and atomically flat interfaces to AFM spintronics~\cite{RN116,RN526,RN603}. One of the first vdW magnets to have been isolated as a monolayer (\ce{CrI3}) exhibits antiferromagnetic behavior in the multilayer regime~\cite{RN7}. The family of vdW AFMs has grown since, with several having been used in spintronic technology demonstrations (including electrical control and terahertz operation)~\cite{RN604,helical_thz}. Semiconducting and insulating vdW AFMs such as \ce{NiI2} are also being studied for their multiferroic properties. The transition metal halide \ce{NiI2} has been known as a helical antiferromagnet for some time~\cite{nii2_foundational}. In the last few years, it has achieved recognition as a vdW type-II multiferroic exhibiting strong interaction between its magnetic and electric polarizations~\cite{nii2_transport,RN606,RN605}. This includes a direct observation of magneto-electric coupling via magnetic modulation of permittivity and electrical polarization~\cite{RN606}. As \ce{NiI2} is a semiconductor with a bandgap on the order of 1 eV, it can be insulating beneath its N\'{e}el temperature and conducts little current without a combination of strong electrostatic gating and contact engineering~\cite{RN605,nii2_transport}. Though the insulating behavior is important to its multiferroic order, it can prove problematic for electrical readouts critical for characterization and devices. Thus, a conducting proxy is needed.

Interfacial magnetotransport phenomena provide a promising route for probing magnetism in electrically insulating materials. At the interface between magnetic and non-magnetic systems, changes in magnetic order can influence charge transport through a variety of interfacial coupling mechanisms, including but not limited to magnetic proximity effects, spin-dependent scattering, and magneto-thermal responses. Such effects enable electrical access to magnetic properties that would otherwise be difficult to probe directly in highly resistive magnets. For example, transport through platinum has been used to identify ferromagnetism in insulating iron yttrium garnet (YIG) thin films~\cite{RN133}, and similar approaches have been demonstrated in platinum/van der Waals \ce{Cr2Ge2Te6} heterostructures~\cite{RN543}. Because these signals originate near the interface, they are most readily detected in materials with strong sensitivity to interfacial perturbations, such as conductors with strong spin–orbit coupling or atomically thin systems. In this context, graphene is a particularly attractive platform due to its two-dimensional nature and pronounced sensitivity to changes at its interface. Indeed, coupling between graphene and the van der Waals antiferromagnet \ce{CrBr3} has been shown to produce substantial modifications of graphene transport~\cite{RN568}. However, analogous interfacial transport studies have not yet been reported for \ce{NiI2}.

Through the fabrication of simple graphene/\ce{NiI2} heterostructures, we demonstrate that graphene magnetotransport can serve as an electrical probe of magnetic phase transitions in multiferroic \ce{NiI2}. The observed magnetotransport features closely track the known magnetic critical temperatures of \ce{NiI2} and aren't known to appear in intrinsic graphene devices, establishing an interfacial origin for the measured signals. These results indicate that changes in the magnetic state of \ce{NiI2} are reflected in the graphene transport response through interfacial coupling. More broadly, our work shows that van der Waals heterostructures incorporating graphene provide a versatile platform for the electrical readout of magnetic phase behavior in insulating multiferroics.

\section{Results and Discussion}\label{sec2}
\begin{figure}[!t]
\centering
\includegraphics[width=\textwidth]{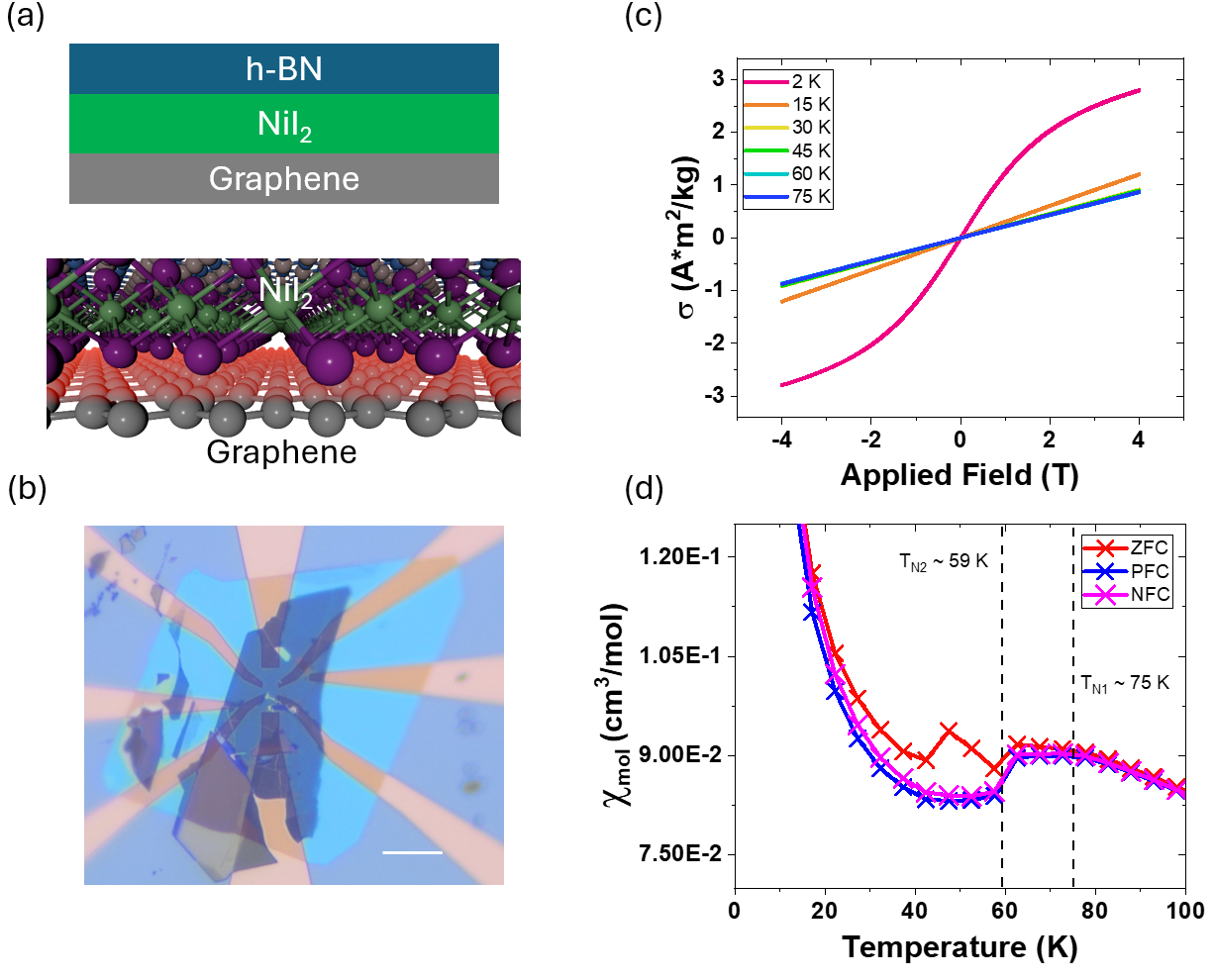}
\caption{\label{fig:f1}(a) Basic diagram of the graphene/\ce{NiI2} heterostructure. (b) Optical image of a Gr/\ce{NiI2}/h-BN heterostructure. This sample is made with monolayer graphene (see Fig.~\ref{fig:nii2_s3}). Scale bar: 10 $\mu$m. (c) Mass-normalized magnetization curves for a field applied in-plane at various temperatures. (d) In-plane susceptibility measurements zoomed in at transition temperatures under zero field cooling (ZFC), negative field cooling (NFC), and positive field cooling (PFC) conditions. Cooling fields were $\pm$ 1 T. For measurements across a larger range, see Fig.~\ref{fig:nii2_s2}.}
\end{figure}

To probe \ce{NiI2}'s magnetic behavior via transport, graphene/\ce{NiI2} bilayers were stacked and placed on pre-patterned Hall bars (see Experimental section). A schematic representation and optical image are shown in Fig.~\ref{fig:f1}a and b, respectively. It bears mentioning that even compared to other vdW magnetic materials, \ce{NiI2} is exceptionally volatile. Indeed, a considerable portion of ref.~\citenum{nii2_transport}'s work was dedicated to a discussion of how to properly preserve \ce{NiI2} heterostructures. To mitigate degradation during fabrication, flakes were transferred in a nitrogen glovebox (see Fig.~\ref{fig:nii2_s1} for more details). A major consequence of this rapid sample degradation is the difficulty in estimating the thickness of the \ce{NiI2} flakes. Given the rapid degradation (even when encapsulated), measurement of thickness after transport characterization via AFM is also not feasible. The absence of AFM thickness metrology is a limitation of the present study because it prevents us from placing each device on a quantitative thickness-dependent \ce{NiI2} phase diagram. However, the central claim of our study does not require assigning an exact flake thickness. Instead, we use the known thickness dependence of \ce{NiI2} as context and compare our transport features to the bulk-like multiferroic transition near 59 K reported in previous work~\cite{nii2_transport}. The optical contrast of the flakes and the appearance of transport features near this temperature are consistent with magnetic behavior close to the bulk limit.

As a point of reference, we first present SQUID measurements of a piece of bulk \ce{NiI2} measured along the in-plane direction. Bidirectional M-H curves measured at various temperatures are shown in Fig.~\ref{fig:f1}c. No significant hysteresis can be seen, consistent with the AFM nature of \ce{NiI2}. At lower temperatures, we see a large non-linear M-H curve emerge. Given the absence of any additional features such as cusps as well as the notable Curie tail visible in Fig.~\ref{fig:f1}d and Fig.~\ref{fig:nii2_s2}a, we attribute this to paramagnetic defects in the crystal, though we note qualitatively similar curves have been attributed to spin-flop transitions in helimagnets previously~\cite{RN606,RN607}. As seen in Fig.~\ref{fig:f1}d, there is a clear kink in the susceptibility at the commonly reported critical temperature T$_{N2}$ (corresponding to the onset of the multiferroic phase) of approximately 59 K~\cite{nii2_transport,nii2_foundational,RN606,RN605}. We also observe a symmetric field-cooling divergence that dissipates at T$_{N2}$, another indicator of AFM ordering. We note that \ce{NiI2} has two commonly reported critical temperatures at 59 K (T$_{N2}$) and 75 K (T$_{N1}$). T$_{N1}$ corresponds to the N\'{e}el temperature of a collinear AFM phase. Our SQUID measurements do not show any obvious features corresponding to T$_{N1}$, though the field cooling divergence closes completely at approximately 75 K and a small uptick near this temperature is apparent in the first derivative of the susceptibility (see Fig.~\ref{fig:nii2_s2}b). These measurements demonstrate the helimagnetic phase transition at the expected temperature and provide a reference to verify whether the observed transport behavior correlates with the critical temperature of the bulk crystal.

\begin{figure}[!t]
\centering
\includegraphics[width=\textwidth]{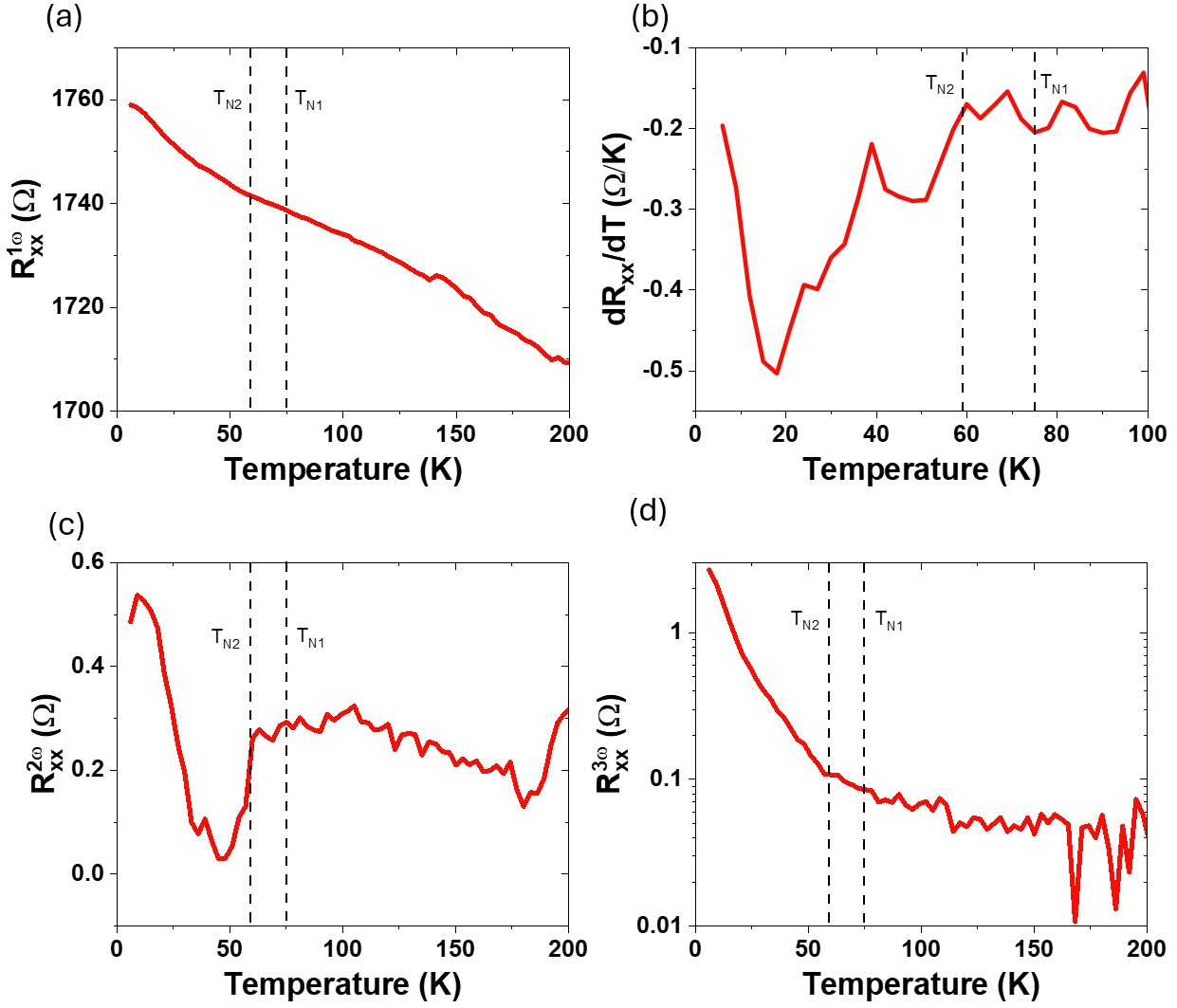}
\caption{\label{fig:f2}(a) Fundamental four-point longitudinal resistance vs. temperature (b) First derivative of data in (a). (c) Second harmonic longitudinal resistance vs. temperature. (d) Third harmonic longitudinal resistance vs. temperature. Note (d) is plotted on a log scale.  In each plot, the dashed lines correspond to critical temperatures.}
\end{figure}

We then proceed to temperature-dependent transport measurements on our crystals. It should be noted that our exfoliated crystals are highly resistive without any gate bias. As such, we do not expect a measurable contribution to resistance due to current shunting through the \ce{NiI2} layer. See Fig.~\ref{fig:nii2_s4} for more details. Upon performing temperature-dependent longitudinal resistance measurements on our heterostructure, we see a monotonic increase in resistance with falling temperature in Fig.~\ref{fig:f2}a. Upon taking a closer look at both the first derivative of the fundamental signal and the second/third harmonics, however, we observe features corresponding to the multiferroic phase transition temperature of \ce{NiI2} as identified by bulk SQUID measurements. Though there is no discernible change in the derivative of the fundamental signal at T$_{N1}$, we see the magnitude of the derivative begin to increase at T$_{N2}$ as shown in Fig.~\ref{fig:f2}b. In the second harmonic longitudinal resistance, we observe a clear drop at 59 K, followed by an increase as the temperature continues to decrease, as shown in Fig.~\ref{fig:f2}c. The third harmonic longitudinal resistance also increases strongly on cooling through the magnetic ordering regime as shown in Fig.~\ref{fig:f2}d. Similar behavior has been observed in other \ce{NiI2}/Gr samples; see Fig.~\ref{fig:nii2_s5}. Notably, control measurements of a monolayer graphene/h-BN sample without \ce{NiI2} but otherwise with the same process (see Fig.~\ref{fig:nii2_s6} for flake information) do not show analogous anomalies at \ce{NiI2}'s critical temperatures, as shown in Fig.~\ref{fig:nii2_s7}. This contrast is especially clear in the second harmonic, whereas the control device also exhibits a sizable third-harmonic background.

As mentioned previously, our \ce{NiI2} crystals are highly resistive. This, combined with the control-device comparison, suggests that the additional structure observed in the nonlinear response originates at the interface rather than from current shunting through the \ce{NiI2} layer. Previous exploration of similar heterostructures using materials like graphene and helical AFMs is limited. At the same time, the control measurements make clear that higher-order transport can also arise in graphene-based devices even without \ce{NiI2}, particularly in the third harmonic. We therefore treat the second- and third-harmonic channels somewhat differently in what follows: the second harmonic provides the clearest evidence for coupling to \ce{NiI2}, whereas the third harmonic likely contains a larger mixture of generic nonlinear transport and thermal effects. Nonlinearities in mono/few-layer graphene have, of course, been studied previously, with optical methods like second harmonic generation (SHG) readily used to demonstrate broken symmetry in strained graphene systems~\cite{graphene_hg_review}. Third harmonic generation does not require broken symmetry, and strong, tunable third harmonic responses have been observed in graphene owing to significant light-matter interaction~\cite{graphene_hg_review,RN608}. Study of higher-order signals in transport is less common, though second harmonic resistance signals in twisted trilayer graphene have been documented, for example~\cite{RN609}. Third harmonic signals are widely used (including in graphene) to study thermal conductivity via the $3\omega$-method~\cite{3w_method,graphene_3w_thermal}. Beyond nonlinear behavior that would appear in graphene alone, both magnetic and ferroelectric systems are rich in such phenomena, with higher-order signals being used to identify ferroelectric ordering and exotic spin textures~\cite{fm_hm_3w,RN487,nonlinear_skyrmion}. Despite this, we are unaware of reports of nonlinear transport in graphene coupled to a magnetic phase transition in an adjacent layer. This motivates a closer look at which parts of the nonlinear response are uniquely modified in the heterostructure via magnetoresistance studies.

\begin{figure}[!t]
\centering
\includegraphics[width=\textwidth]{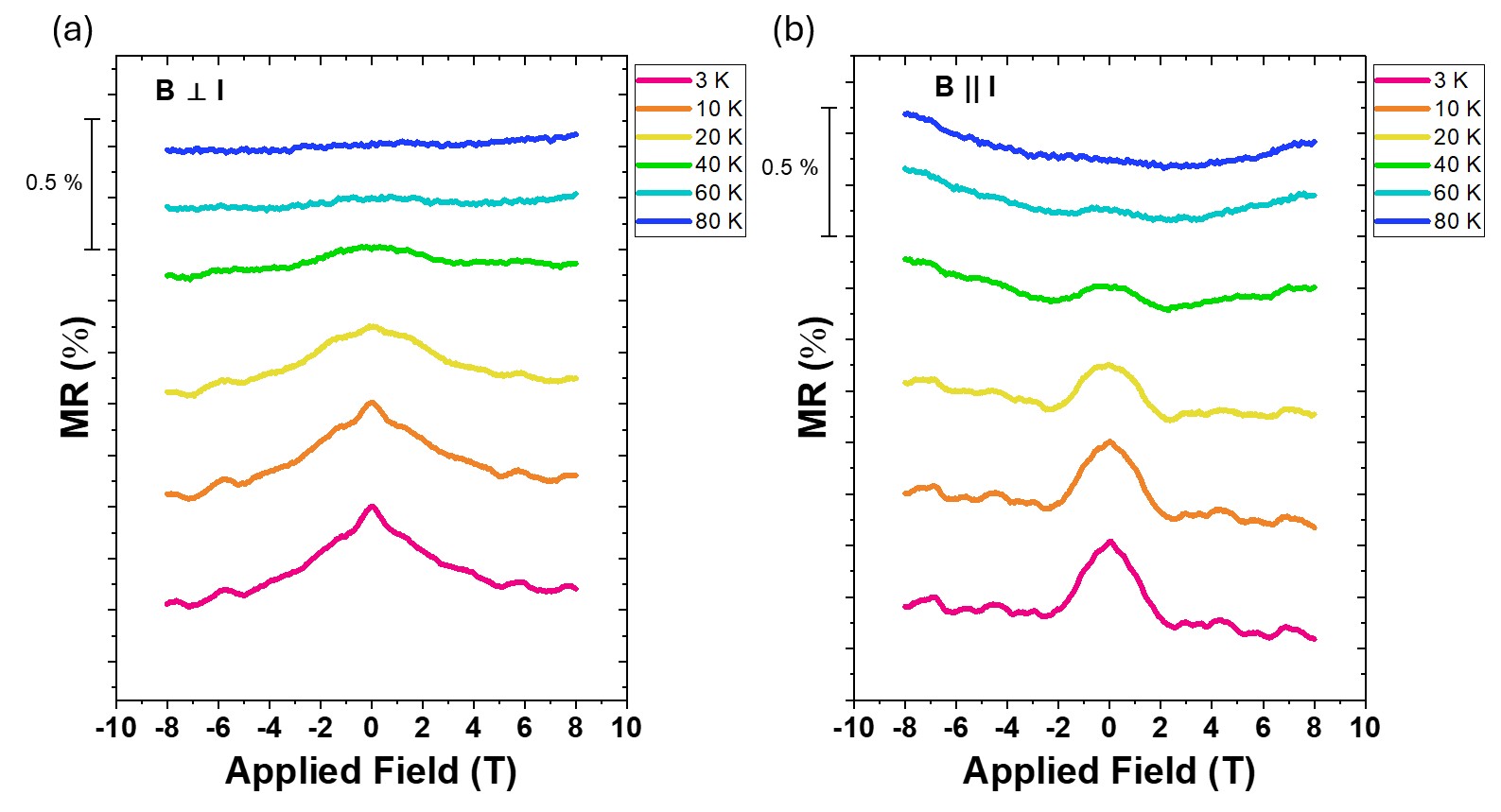}
\caption{\label{fig:f3}(a) Magnetoresistance sweeps for an in-plane field applied perpendicular (90 degrees) to the direction of current flow. (b) Magnetoresistance sweeps for an in-plane field applied parallel (0 degrees) to the direction of current flow. Measurements were conducted with a four-terminal geometry and represent the fundamental (first harmonic) response.}
\end{figure}

To probe magnetotransport, we primarily study the longitudinal magnetoresistance (calculated as $MR = \frac{R_{xx}(B)-R_{xx}(0)}{R_{xx}(0)}$, where $R_{xx}(B)$ is the measured resistance at a certain field and $R_{xx}(0)$ is the measured resistance at zero field) of the fabricated heterostructures under various field and temperature conditions. Initial measurements of the Hall response in a Gr/\ce{NiI2} heterostructure only showed linear behavior expected for graphene, as seen in Fig.~\ref{fig:nii2_s8}. While anomalous Hall signals are sometimes reported in graphene/magnet heterostructures, none were found here. Additionally, out-of-plane longitudinal magnetoresistance measurements yielded parabolic curves with little temperature dependence, consistent with the geometric magnetoresistance that is well-documented in graphene (see Fig.~\ref{fig:nii2_s9}a). For an in-plane field, we would not expect to see any magnetoresistance in a graphene system (except for a small parabolic component stemming from a misalignment of the applied field). Instead, we observe large peak shapes centered about zero field as can be seen in Figs.~\ref{fig:f3}a and~\ref{fig:f3}b. This first-harmonic in-plane magnetoresistance is the clearest direct magnetotransport signature in the present devices. Notably, these peak features reduce in prominence as the temperature rises and disappear above the multiferroic transition temperature of \ce{NiI2} (see Fig.~\ref{fig:nii2_s11}). There is also a clear anisotropy in the measured signals, with the peak being sharper for a field applied parallel to the current direction. Such peak-shaped features have been observed in several additional heterostructures (see Figs.~\ref{fig:nii2_s9} and~\ref{fig:nii2_s10}). Taken together, the strong correlation with \ce{NiI2}’s magnetic phase transition and the inconsistency with known graphene magnetoresistance mechanisms strongly implicate an interfacial origin tied to the magnetic ordering of \ce{NiI2}. As with the resistance vs. temperature measurements, the graphene control device did not exhibit any of this behavior (see Fig.~\ref{fig:nii2_s12}). It only exhibited a small parabolic magnetoresistance, likely geometric magnetoresistance originating from a slight misalignment of the magnetic field. Though it was fabricated in the same way, the graphene-only control device cannot be used as a one-to-one background subtraction: graphene geometry, disorder, and electrostatic environment, including the absence of adjacent multiferroic \ce{NiI2}, can all affect the graphene response. Its role is instead to show that similarly measured monolayer graphene/h-BN devices do not exhibit the large low-field in-plane magnetoresistance peaks or sharp second-harmonic anomalies seen in the \ce{NiI2}/graphene heterostructure, features that are not expected for ordinary graphene/h-BN magnetotransport.

To interpret this behavior, we consider a phenomenological interfacial-scattering picture in which the magnetic state of \ce{NiI2} modifies transport in the adjacent graphene layer. In conductor/insulator heterostructures, magnetic ordering in an insulating layer can influence charge transport in a neighboring conductor through magnetic proximity effects, spin-dependent scattering, and related interfacial mechanisms~\cite{RN133,RN543,RN568,cgt_graphene_prox}. For a helical antiferromagnet such as \ce{NiI2}, an applied magnetic field can distort the spin spiral, reconfigure helical domains, or drive evolution toward fan-like or partially polarized states~\cite{helical_afm_theory,RN610,helimagnet_phase_transition}. Such field-induced changes would be expected to modify spin-dependent interfacial scattering and can qualitatively account for a low-field magnetoresistance feature that weakens at higher fields. Because the crystallographic orientation and helical-domain population of the exfoliated \ce{NiI2} flake are not known, we do not assign the two measured field geometries to specific orientations relative to the helical propagation vector. Instead, the angular dependence is taken as evidence that the interfacial transport response is anisotropic and coupled to field-driven changes in the magnetic state of \ce{NiI2}. Transport alone, however, cannot uniquely determine the microscopic mechanism.

\begin{figure}[!t]
\centering
\includegraphics[width=\textwidth]{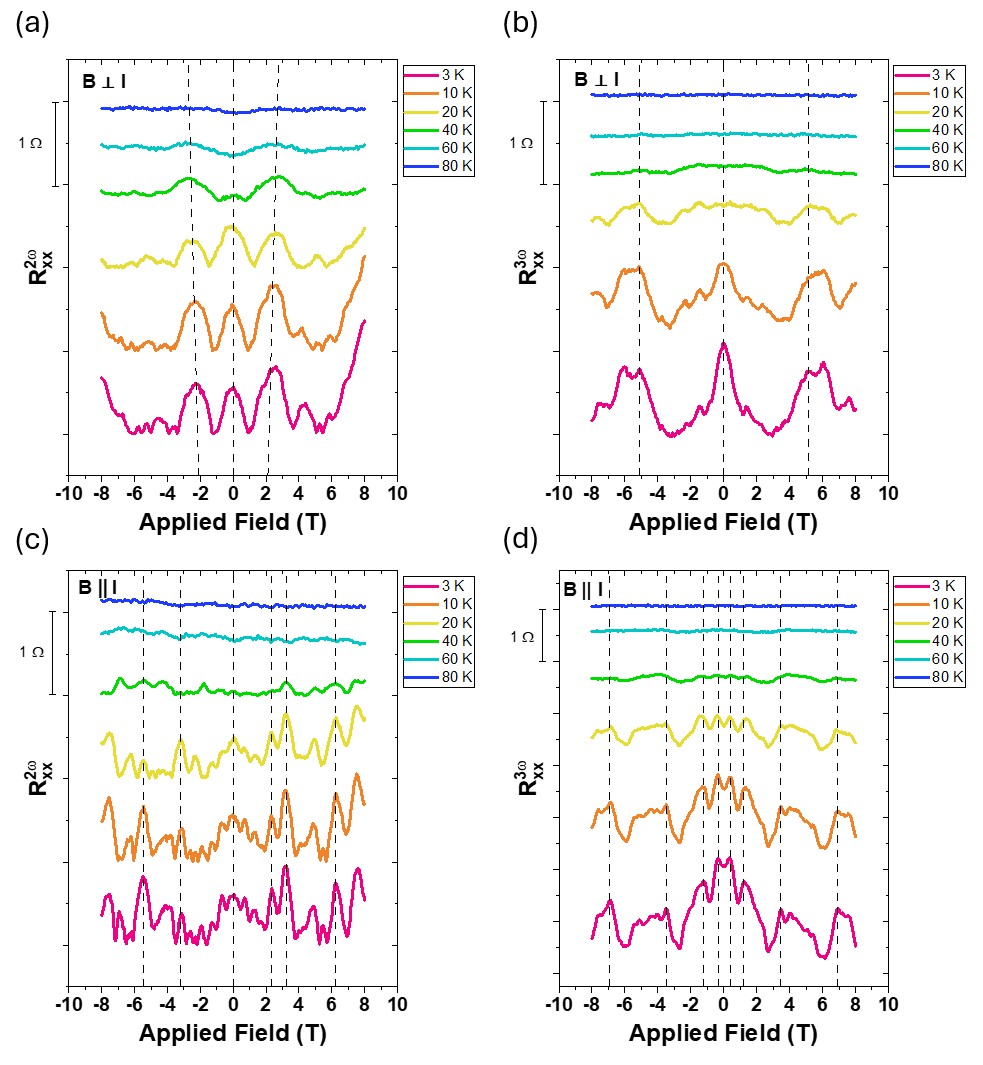}
\caption{\label{fig:f4}(a) Second and (b) third harmonic resistance of MR sweeps shown in Fig.~\ref{fig:f3}a (field perpendicular to current). (c) Second and (d) third harmonic resistance of MR sweeps shown in Fig.~\ref{fig:f3}b (field parallel to current). The dashed lines in each figure highlight peaks that appear across lower temperature measurements.}
\end{figure}

Next, we turn our attention to nonlinearities in the magnetoresistance signals. As shown in Fig.~\ref{fig:f2}, strong nonlinear transport behavior emerges coincident with the onset of magnetic ordering in \ce{NiI2}. Given this link to the magnetic state of \ce{NiI2}, we might also expect to see magnetoresistive behavior in the higher-order signals. Figure~\ref{fig:f4} shows the measured second- and third-harmonic resistances of the in-plane magnetoresistance signals presented in Fig.~\ref{fig:f3}. In the heterostructure, the second harmonic develops a series of pronounced, reproducible peak-like features. The response also depends on the direction of the applied magnetic field, consistent with anisotropic behavior. When compared to nonlinear magnetotransport in the graphene control device (Fig.~\ref{fig:nii2_s12}b and c), the clearest contrast appears in the second harmonic: the control device shows no analogous peak-like structures and only a comparatively smooth increase at higher fields. This makes the $2\omega$ response the most direct nonlinear transport signature associated with the presence of \ce{NiI2}. The peak-like structures in the \ce{NiI2}/graphene heterostructure may therefore reflect enhanced sensitivity of the nonlinear channel to field-driven changes in the helical magnetic state, such as spin-spiral distortion, domain reconfiguration, or partial unwinding. We do not assign individual peaks to specific magnetic phases; rather, their reproducibility, anisotropy, and absence in the graphene control device indicate an additional nonlinear contribution associated with \ce{NiI2}.

To further analyze the harmonic magnetotransport signals, we consider possible physical origins of the observed nonlinearities. Second-harmonic ($2\omega$) signals are commonly associated with broken symmetries, which in the present heterostructure may arise from both the graphene/\ce{NiI2} interface itself and the emergence of multiferroic order in \ce{NiI2}~\cite{RN487,RN611,shear_mediated}. In addition to a finite background $2\omega$ response, likely associated with interfacial asymmetry, the pronounced change in the second-harmonic signal below T$_{N2}$ and its much richer field dependence relative to the control device indicate an additional contribution that emerges with magnetic ordering in \ce{NiI2}. The non-monotonic temperature dependence suggests that multiple interfacial mechanisms contribute to the $2\omega$ response, with distinct temperature and field dependences. Possible contributors include magneto-thermoelectric effects or other interfacial nonlinear processes, though the present data do not allow us to isolate a unique microscopic mechanism. For the purposes of this work, the key point is that the $2\omega$ response changes sharply as \ce{NiI2} enters its ordered phases and is far more structured than in the graphene-only control.

The third-harmonic ($3\omega$) response requires greater caution. Third-harmonic signals are known to be highly sensitive to Joule heating (and are in fact used to characterize thermal properties~\cite{dames2013measuring}) as well as to other nonlinear responses, and the graphene control device likewise exhibits a strong $3\omega$ signal. We therefore interpret the third harmonic as a mixed signal containing a substantial background from generic nonlinear transport, likely including magneto-thermal effects. Even so, the \ce{NiI2}/Gr heterostructure shows a different temperature dependence from the control, with changes beginning near $T_{N1}$ and strengthening below T$_{N2}$. This suggests that magnetic order in \ce{NiI2} modulates the background nonlinear response, rather than generating a wholly distinct third-harmonic signal by itself. Given the limited literature on low-frequency nonlinear transport in such heterostructures, we refrain from assigning a unique microscopic origin here. Overall, the control comparison indicates that the clearest \ce{NiI2}-specific nonlinear signature resides in the second harmonic, while the third harmonic remains consistent with an additional magnetic contribution superimposed on a broader background.

Additional current-dependent measurements for both the heterostructure and the graphene-only control are provided in Figs.~\ref{fig:nii2_s14} and~\ref{fig:nii2_s15}. The heterostructure shows much stronger bias dependence and more non-trivial harmonic scaling than the control device, indicating that bias-induced effects are strongly device dependent. These measurements support the view that self-heating and related bias effects influence the nonlinear transport, but they do not by themselves isolate a unique microscopic origin.

\section{Conclusion}\label{sec13}

We have demonstrated that first-harmonic and nonlinear magnetotransport in graphene/\ce{NiI2} heterostructures enables sensitive electrical probing of magnetic phase transitions in a helical antiferromagnetic multiferroic. The central signature is a pronounced, anisotropic in-plane first-harmonic magnetoresistance response that is suppressed above the \ce{NiI2} multiferroic transition and is absent from a monolayer graphene control device. Comparison to this control device shows that the clearest \ce{NiI2}-specific nonlinear signature appears in the second harmonic, while the third harmonic is better interpreted as a mixed response containing a substantial generic nonlinear background that is nevertheless modified by the magnetic state of \ce{NiI2}. Taken together with the supplementary current-dependent measurements on both the heterostructure and graphene-only control, our results highlight how changes in complex magnetic order can be transduced into measurable graphene transport signatures. The present work establishes an experimental route for electrically accessing magnetic-state-dependent interfacial responses in graphene/\ce{NiI2} heterostructures, while leaving the microscopic origin of the observed magnetotransport signatures as an open question for future theoretical and materials-specific studies. These findings point toward opportunities for low-power electrical readout of antiferromagnetic and multiferroic states and establish graphene-based heterostructures as a versatile platform for integrating insulating van der Waals magnets into spintronic devices.

\section{Experimental Methods}\label{sec11}

\textit{Sample Fabrication}: Ti/Au Hall bars were patterned onto 285 nm \ce{SiO2}/Si wafers using either photolithography or electron beam lithography. Blank Hall bars were then taken into a nitrogen glovebox where the \ce{NiI2} (crystals obtained from 2D Semiconductors) was exfoliated. Separately, graphene flakes were exfoliated under ambient conditions, and substrates containing suitable candidates for transfer (as determined by optical appearance) were brought into the glovebox. Using a transfer stage in the glovebox, exfoliated \ce{NiI2} flakes were picked up using polycarbonate stamps~\cite{RN201}. Upon successful pickup of \ce{NiI2}, the selected flake was then used to pick up an exfoliated graphene flake. The newly formed bilayer was then stamped onto the pre-patterned Hall bar such that the graphene sits atop the Au electrodes to form electrical contacts. 

\textit{Measurements}: SQUID measurements in Fig.~\ref{fig:f1}c and d were performed with a Quantum Design MPMS3 VSM. Field ranges up to 1000 Oe were used to measure susceptibility. Magnetoresistance characterizations were carried out using the ETO option of a Quantum Design PPMS. Unless otherwise stated, excitations were 10 $\mu$A at 3 Hz. Harmonic signals were obtained from the ETO-reported second- and third-harmonic outputs. According to the ETO measurement convention, these correspond to in-phase harmonic voltage amplitudes; when reported in dB, the harmonic levels are referenced to the fundamental voltage response. We converted the reported harmonic levels to voltages using the measured first-harmonic signal. Additional electrical characterization, such as probing the resistance of the \ce{NiI2} itself, was performed in a Lakeshore probe station with a B1500A semiconductor analyzer. 

\section*{Funding}

The authors would like to acknowledge the support from the University of Illinois at Urbana-Champaign through the Campus Research Board Award, and from the Center for Advanced Semiconductor Chips with Accelerated Performance (ASAP) under NSF Grant no. EEC-2231625.

\section*{Acknowledgments}

The authors acknowledge the use of facilities and instrumentation supported by NSF through the University of Illinois at Urbana-Champaign Materials Research Science and Engineering Center DMR-2309037.

\section*{Competing Interests}

The authors declare no financial or non-financial competing interests.

\section*{Author Contributions}

W.J. and S.T.C. conceived the idea of graphene/\ce{NiI2} heterostructures. S.T.C. carried out fabrication and measurements of the \ce{NiI2}-bearing devices. X.X. fabricated and performed characterization of the four-point graphene control device. S.T.C., X.X., and W.J. all contributed to the manuscript.

\clearpage

\begin{center}
\begin{Huge}
    \textbf{Supplementary Information}
\end{Huge}
\end{center}
\clearpage

\setcounter{figure}{0}
\renewcommand{\figurename}{Figure}
\renewcommand{\thefigure}{S\arabic{figure}}
\renewcommand{\theHfigure}{S\arabic{figure}}

Collection of images showing sample degradation under various conditions. As previously reported in Ref.~\citenum{nii2_transport}, \ce{NiI2} degrades rapidly unless special precautions are taken. Given our fabrication and measurement considerations, we opted to leave samples covered in h-BN and/or polycarbonate to slow (but not completely halt) degradation. Measurements were always conducted soon after fabrication to mitigate the effects of degradation on the results. This would not protect the \ce{NiI2} indefinitely, and within the span of a couple of weeks, the flake significantly degrades. Capping the bilayer with hexagonal boron nitride and, in turn, covering the full stack with polycarbonate was also tried (as in the case of the sample in Fig.~\ref{fig:f1}b), though it did not yield significantly better results. Given the absence of an AFM system in our glovebox, these difficulties made it impossible to precisely measure the thickness of the \ce{NiI2} flakes used in our heterostructures. However, based on the thickness-dependent appearance and behavior reported in~\citenum{nii2_transport}, we believe our flakes are thick enough to have magnetic properties close to bulk. This comparison is therefore used only to support a bulk-like magnetic-transition assignment.

\begin{figure}[hbt!]
\centering
\includegraphics[width=\columnwidth]{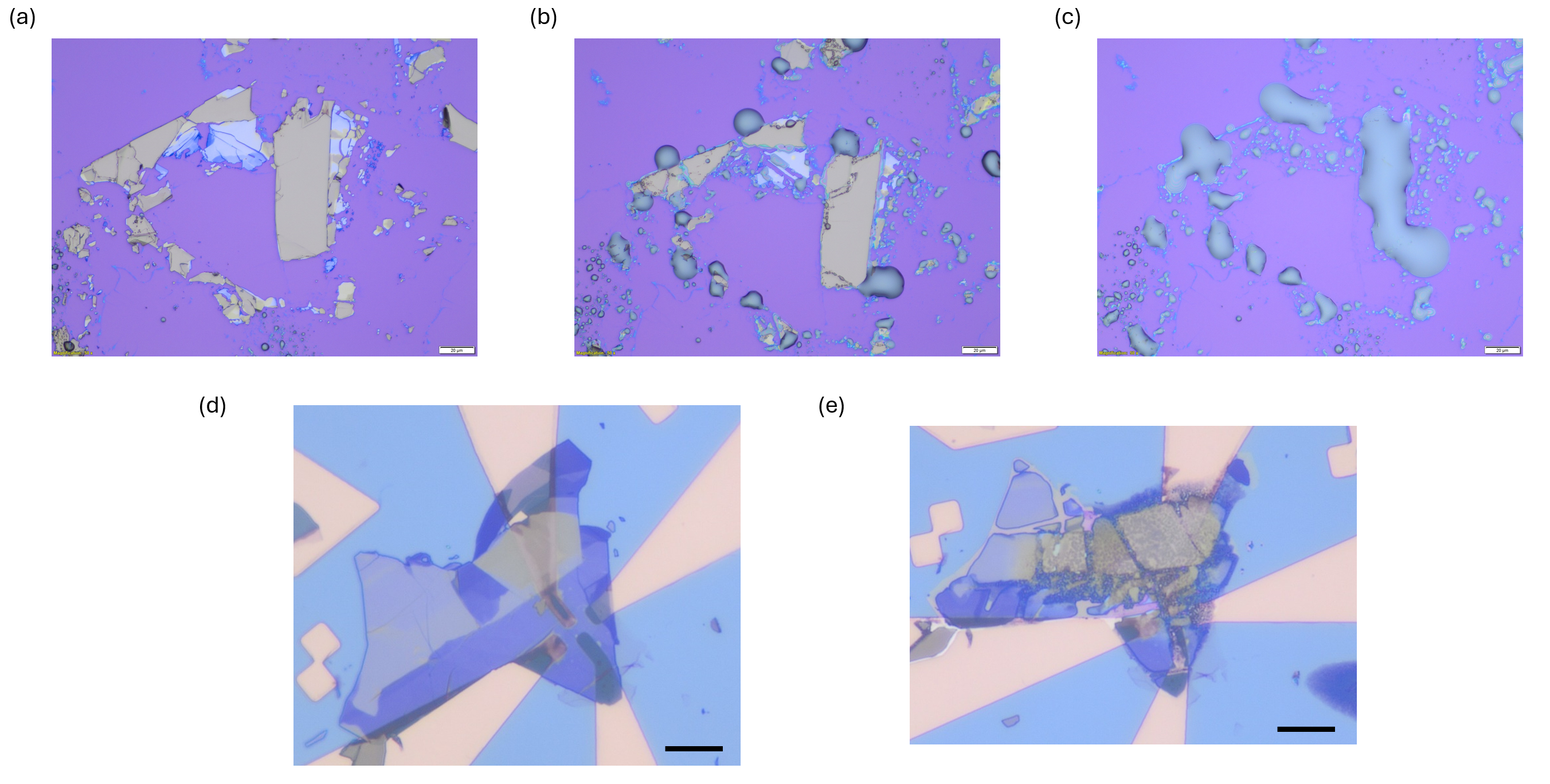}
\caption{\label{fig:nii2_s1}(a)-(c) Optical images of \ce{NiI2} flakes degrading under exposure to ambient conditions. Images are taken immediately after glovebox removal, five hours after glovebox removal, and three days after glovebox removal. (d), (e) Optical images of a Gr/\ce{NiI2} heterostructure (sample 2) capped only with polycarbonate, immediately after fabrication and after ten days, respectively.}
\end{figure}
\clearpage

\begin{figure}[hbt!]
\centering
\includegraphics[width=\columnwidth]{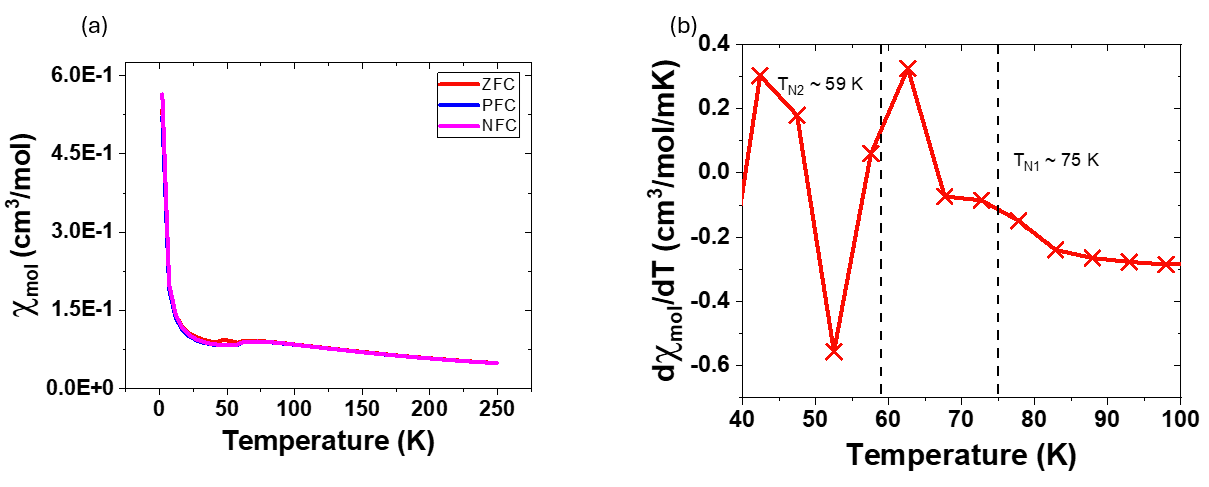}
\caption{\label{fig:nii2_s2}(a) Measured in-plane susceptibility across the full temperature range for our crystals. (b) First derivative of susceptibility data for zero field cooling around critical temperatures.}
\end{figure}

Full range of SQUID results and look at the first derivative data near critical points. Instrument time constraints prevented a high-resolution scan of the region; however, an uptick of the susceptibility around the first critical point T$_{N1}$ can be seen in addition to the oscillations around T$_{N2}$. As can be seen in Fig.~\ref{fig:nii2_s2}a, there is a large increase in susceptibility as the temperature approaches 0 K, likely a Curie tail from paramagnetic defects.

\clearpage

Optical image of the graphene flake used in the main text's heterostructure and its extracted contrast profile. We measure a contrast of approximately 9.6\%, indicating the flake is likely a monolayer~\cite{RN352}.

\begin{figure}[hbt!]
\centering
\includegraphics[width=\columnwidth]{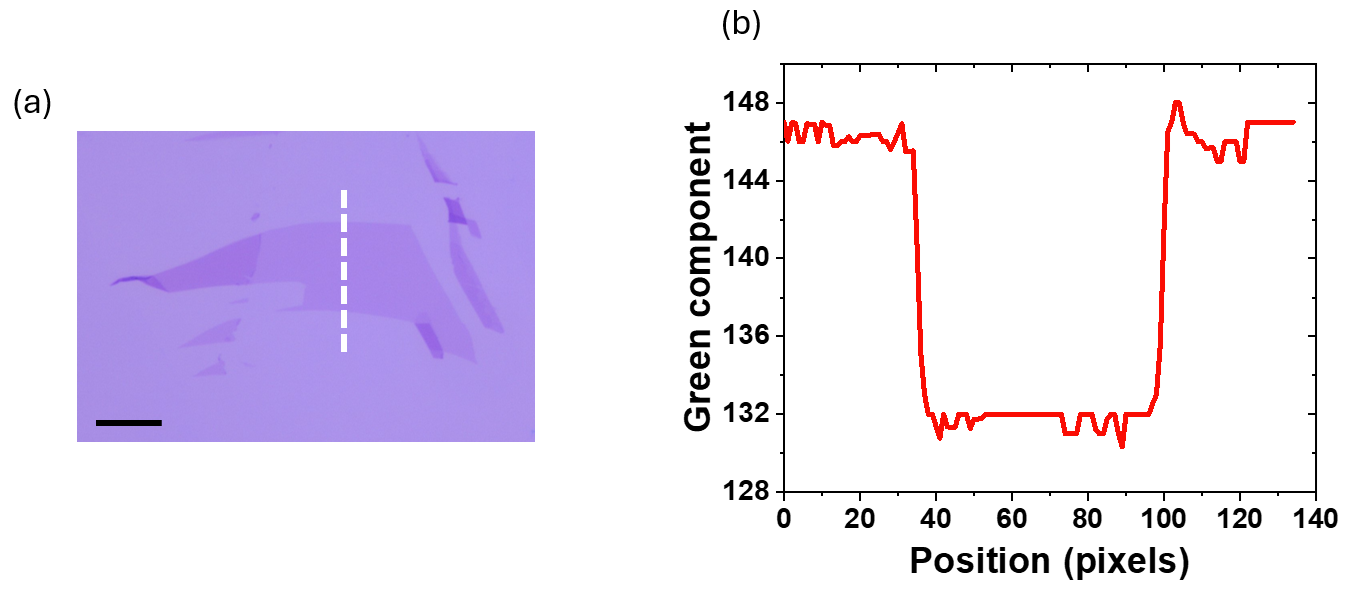}
\caption{\label{fig:nii2_s3}(a) Optical image of graphene used in the main text's heterostructure. Scale bar: 10 $\mu$m. (b) Profile of green component along white dotted line in (a) used to extract contrast.}
\end{figure}
\clearpage

\begin{figure}[hbt!]
\centering
\includegraphics[width=\columnwidth]{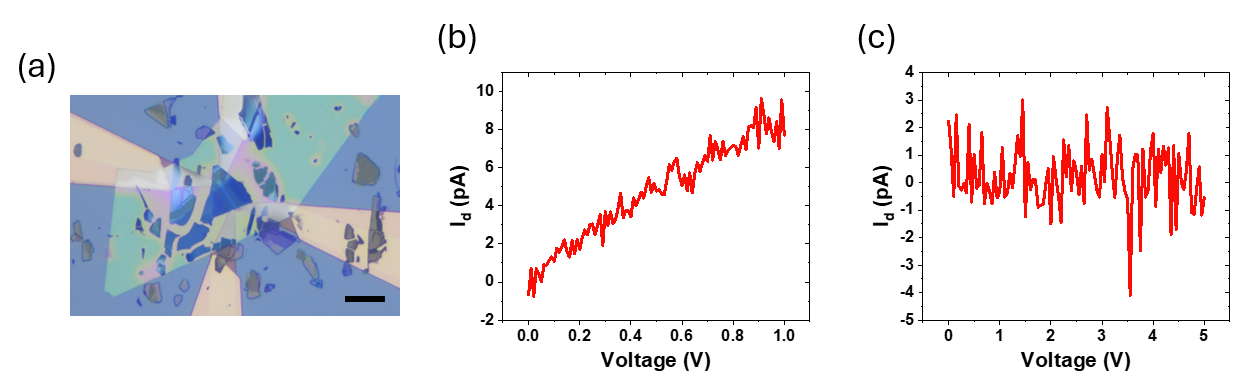}
\caption{\label{fig:nii2_s4}(a) Optical image of \ce{NiI2}-only sample made with two platinum contacts. Scale bar: 10 $\mu$m. (b) Two-point IV measurement at room temperature under zero gate bias. (c) Two-point IV measurement at $\sim 8$ K under zero gate bias.}
\end{figure}

Our \ce{NiI2} crystals are highly insulating under zero gate bias, such that we do not expect any significant current shunting through them in the Gr/\ce{NiI2} heterostructures. As an example, we present two-point resistance data from one sample taken in a probe station. This sample has two platinum contacts separated by 1.5 $\mu$m. It was originally intended for inverse spin Hall effect measurements, though that experiment did not provide significant results. As can be seen in Fig.~\ref{fig:nii2_s4}b, the sample is highly resistive at room temperature, with a two-point resistance on the order of 100 G$\Omega$. Shown in Fig.~\ref{fig:nii2_s4}c, at low temperature (probe station's base temperature; $\sim 8$ K), we cannot get a clean signal, suggesting that the resistance is much greater as would be expected for a semiconductor.

\clearpage

\begin{figure}[hbt!]
\centering
\includegraphics[width=\columnwidth]{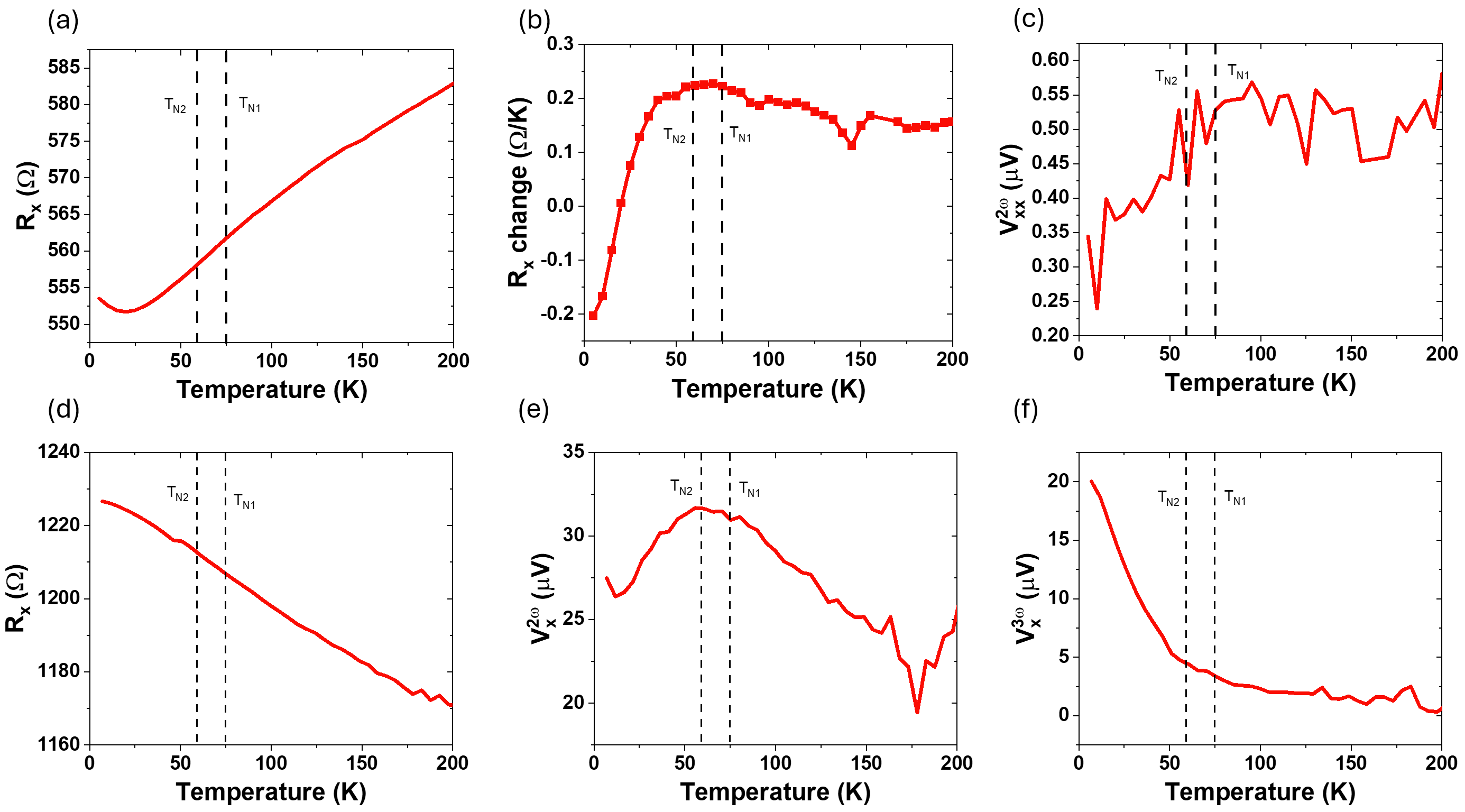}
\caption{\label{fig:nii2_s5}(a) Resistance vs. temperature for sample 2. (b) Derivative of data in (a). (c) Second harmonic data in (a). (d) Resistance vs. temperature for sample 3. (e) Second harmonic of data in (d). (f) Third harmonic of data in (d).}
\end{figure}

\begin{figure}[hbt!]
\centering
\includegraphics[width=\columnwidth]{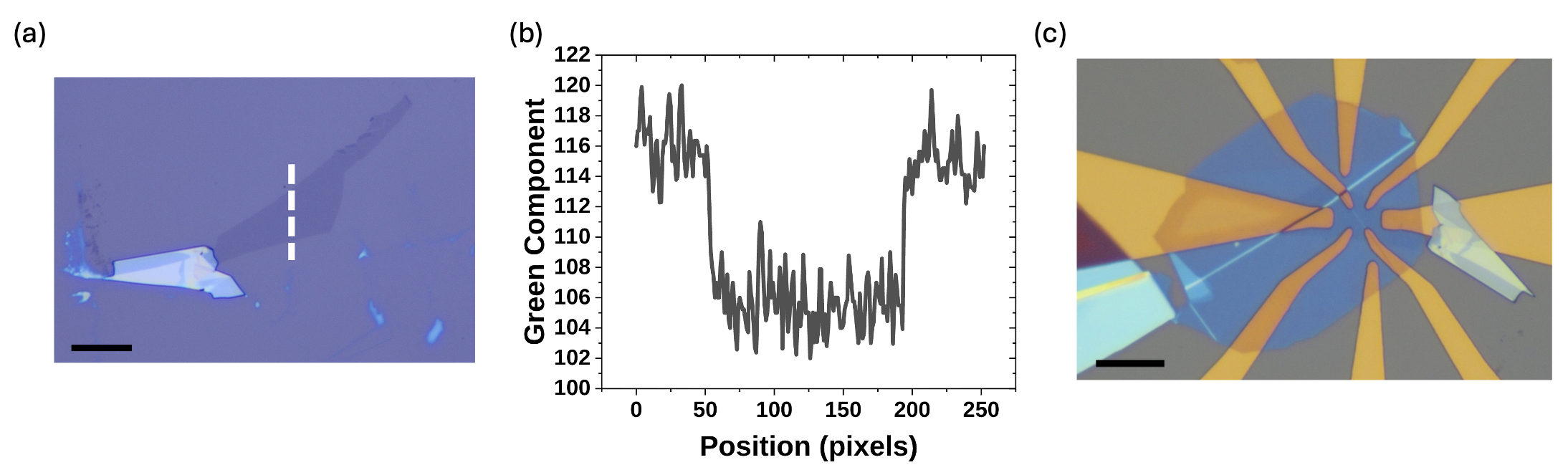}
\caption{\label{fig:nii2_s6}(a) Optical image of the control device flake before transfer. Scale bar: 10 $\mu$m. (b) Profile of green component along white dashed line in (a). (c) Optical image of completed device. Scale bar: 10 $\mu$m.}
\end{figure}

\begin{figure}[hbt!]
\centering
\includegraphics[width=\columnwidth]{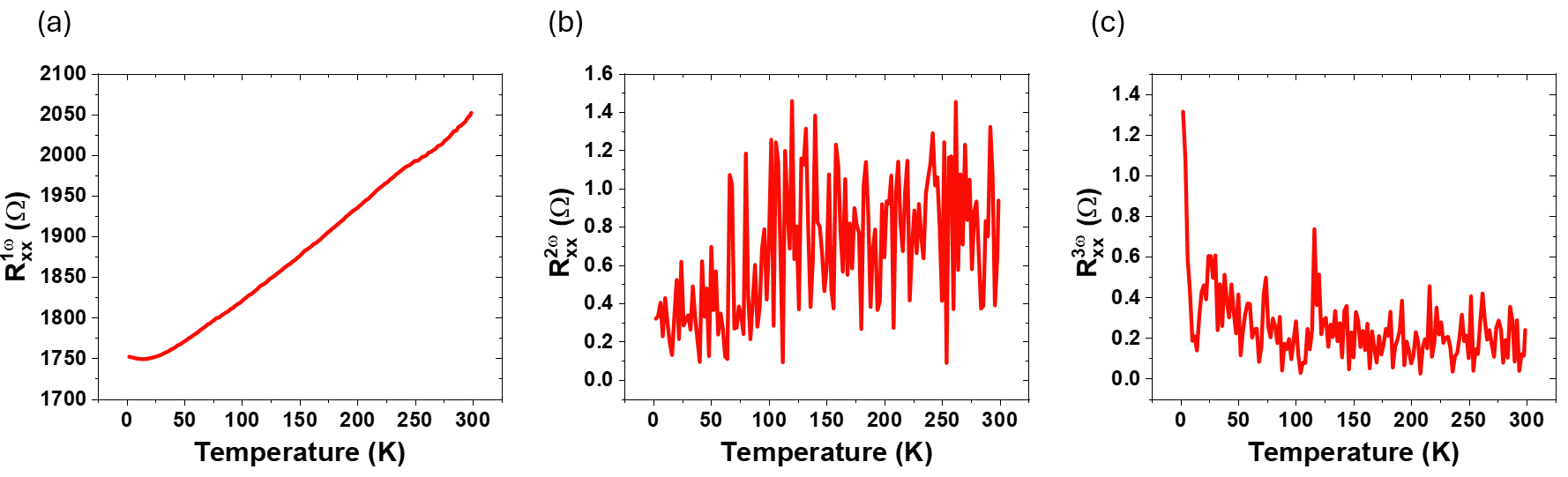}
\caption{\label{fig:nii2_s7}(a) Resistance vs. temperature for graphene control sample. (b) Second harmonic of data in (a). (c) Third harmonic of data in (a).}
\end{figure}

Additional temperature-dependent resistance data from two more heterostructures (sample 2 and sample 3). Measurements used a 10 $\mu$A drive current. Note that these samples were made with four-point Hall bars and thus a two-point resistance is measured. This means that contact resistance may also influence the results. Interestingly, sample 2's resistance as reported in Fig.~\ref{fig:nii2_s5}a generally decreases with temperature until about 20 K, in contrast to other devices, possibly due to the influence of the contacts. Plotting the derivative of this data vs. temperature, we can see changes at \ce{NiI2}'s critical points in Fig.~\ref{fig:nii2_s5}b. The phase transition also appears in the second harmonic data in Fig.~\ref{fig:nii2_s5}a. Third harmonic data for this sample did not have any discernible trends. Compared to other samples, this one had a much wider graphene channel and, consequently, a lower resistance. It's possible that, in tandem with contact effects, there was not enough heating in the device to generate higher-order signals. Resistance vs. temperature data for sample 3, shown in Fig.~\ref{fig:nii2_s5}d, follows the typical trend and is also measured with only two points. It also shows a strong temperature dependence in the higher-order signals, as can be seen in Figs.~\ref{fig:nii2_s5}e and f. Interestingly, the second harmonic voltage for this sample is much larger than the third, in contrast with the data from sample 1 in the main text. Again, nonlinearities stemming from the contact regions could explain the difference.

\clearpage

Optical images of the graphene flake used in a four-point control sample as well as an optical image of the completed device. We measure an optical contrast of approximately 8.5\%, again consistent with a monolayer~\cite{RN352}. This device was made in the same fashion as the \ce{NiI2} devices.

\clearpage

Temperature-dependent four-point resistance data from our control graphene device with graphene in contact with h-BN. The flake is similar in size to the main device, and a drive current of 10 $\mu$A was also used for measurements. Notably, there are no obvious changes in the transport trends near \ce{NiI2}'s critical temperature like there are with the samples incorporating it into the heterostructure. We also note that the general trend in the resistance is opposite that of the main \ce{NiI2} device; e.g., it decreases instead of increases with falling temperature. Given the sensitivity of graphene to its surroundings there are many possible reasons for this. One possible cause is differences in doping from the adjacent flakes, though it is difficult to say definitively with only transport measurements. 

\clearpage

\begin{figure}[hbt!]
\centering
\includegraphics[width=\columnwidth]{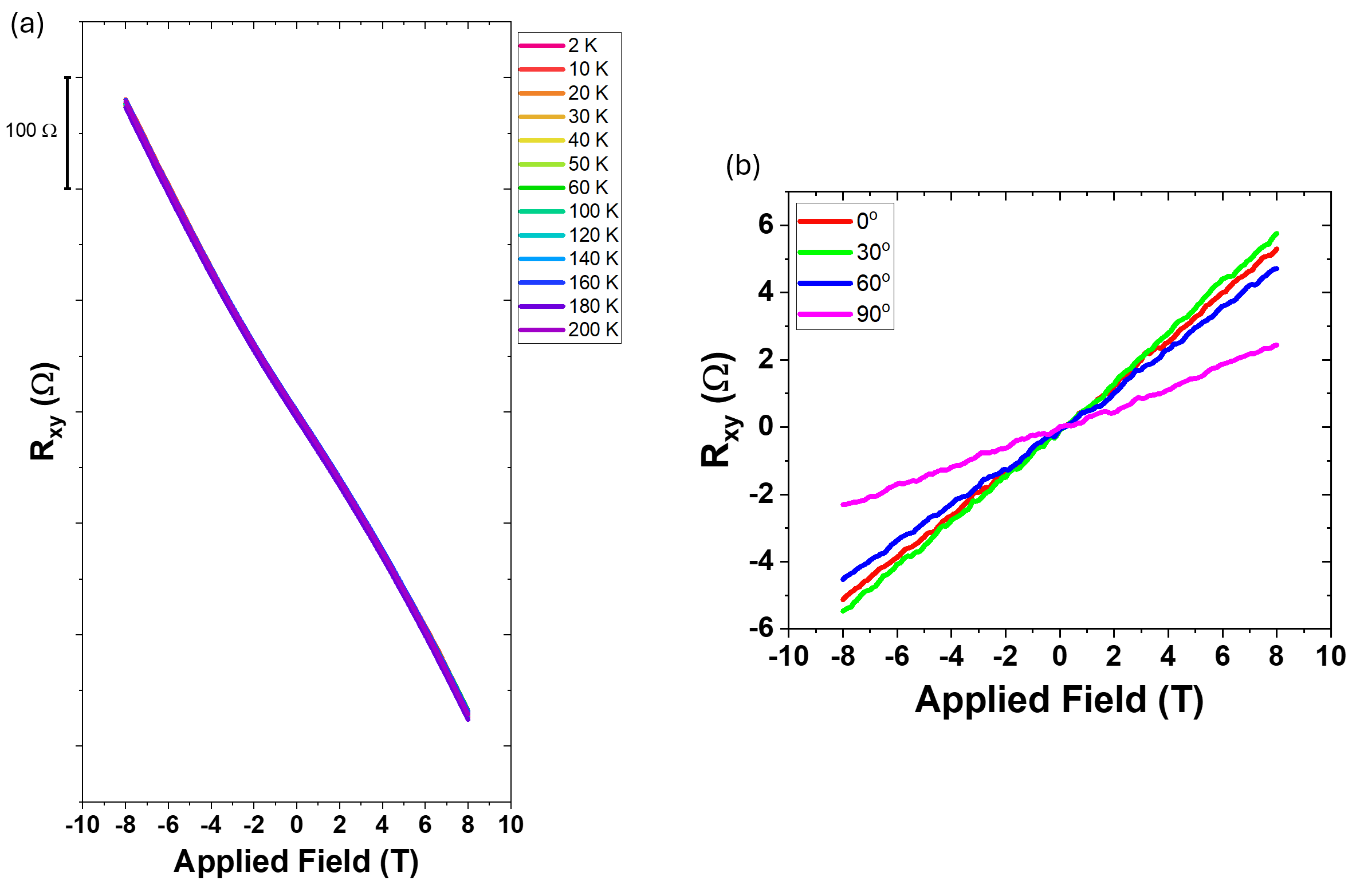}
\caption{\label{fig:nii2_s8}(a) Out-of-plane Hall effect measurements for sample 2 at various temperatures. (b) In-plane Hall effect data for sample 4 measured at various angles at a temperature of 2 K. Angles are with respect to the direction of current flow; e.g., 0 degrees is parallel to the current.}
\end{figure}

\begin{figure}[hbt!]
\centering
\includegraphics[width=\columnwidth]{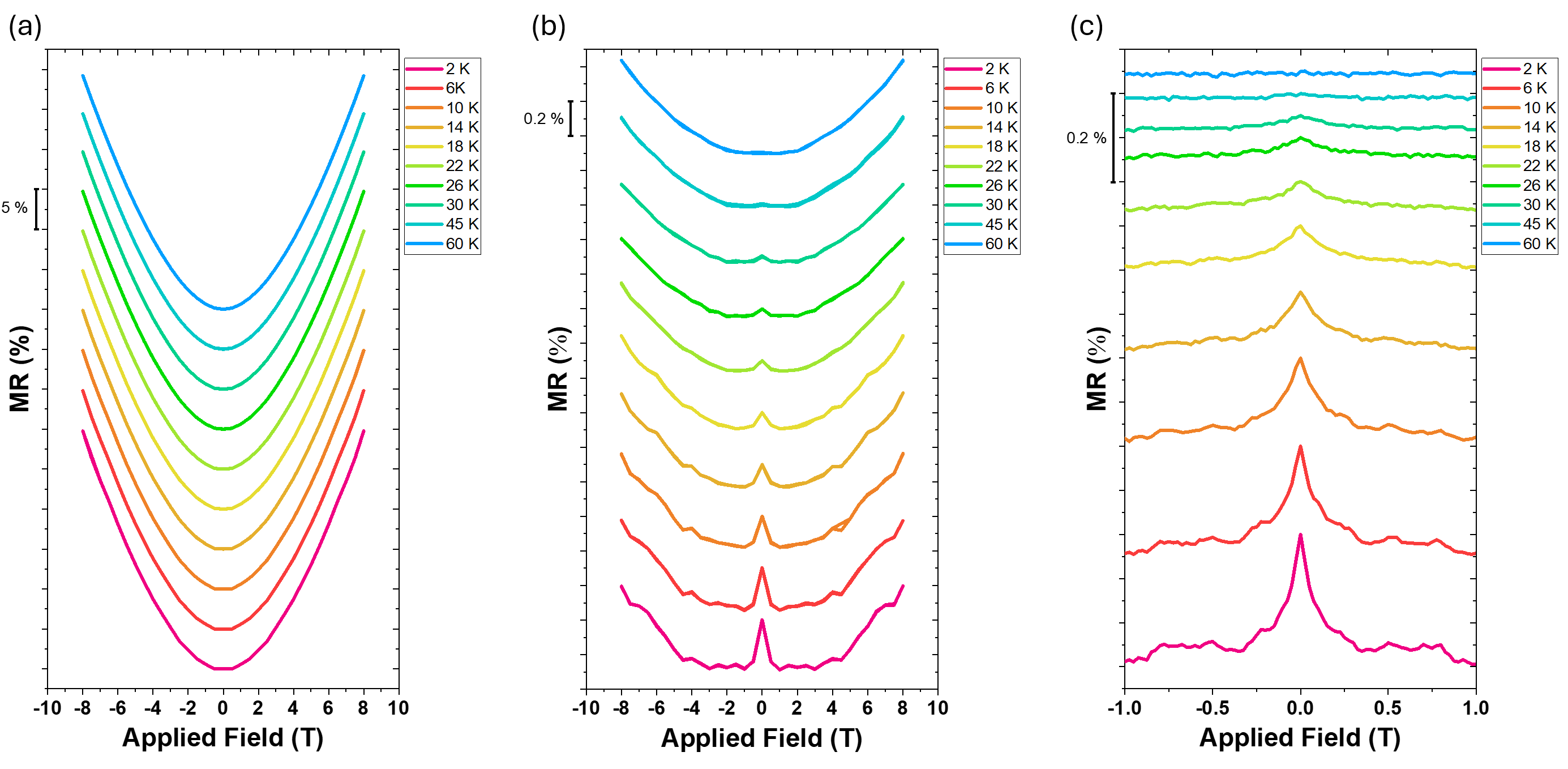}
\caption{\label{fig:nii2_s9}(a) Out-of-plane magnetoresistance data for sample 2 at various temperatures. (b) In-plane (parallel to current) magnetoresistance data for sample 2 at various temperatures. (c) Higher resolution field scan of in-plane magnetoresistance across a smaller range. Both (a) and (b) used full forwards and back hysteresis sweeps, whereas (c) only goes in one direction.}
\end{figure}

\begin{figure}[hbt!]
\centering
\includegraphics[width=\columnwidth]{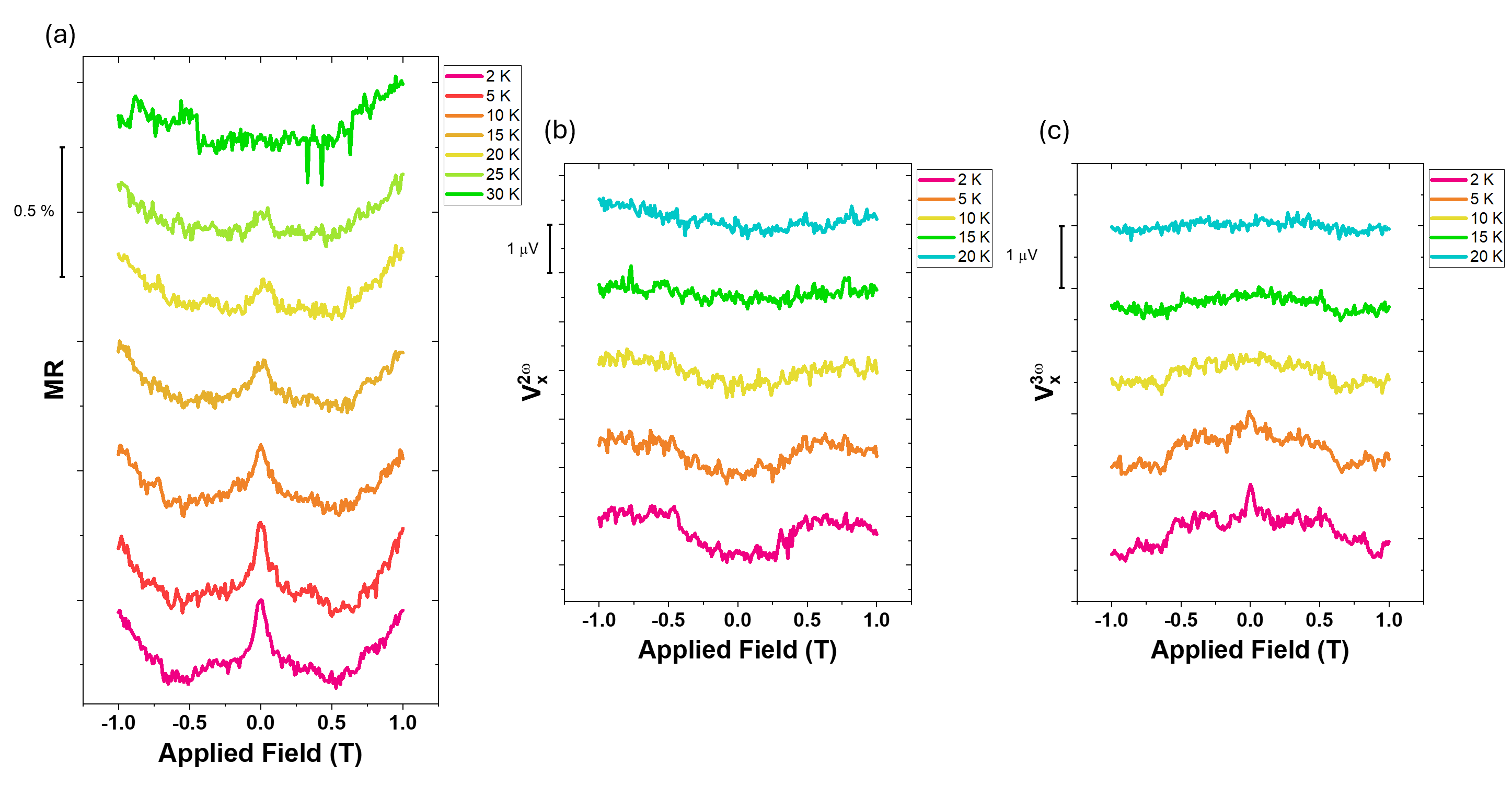}
\caption{\label{fig:nii2_s10}(a) In-plane (parallel to current) fundamental magnetoresistance data for sample 3. (b) Second and (c) third harmonic voltage of magnetoresistance data.}
\end{figure}

\begin{figure}[hbt!]
\centering
\includegraphics[width=\columnwidth]{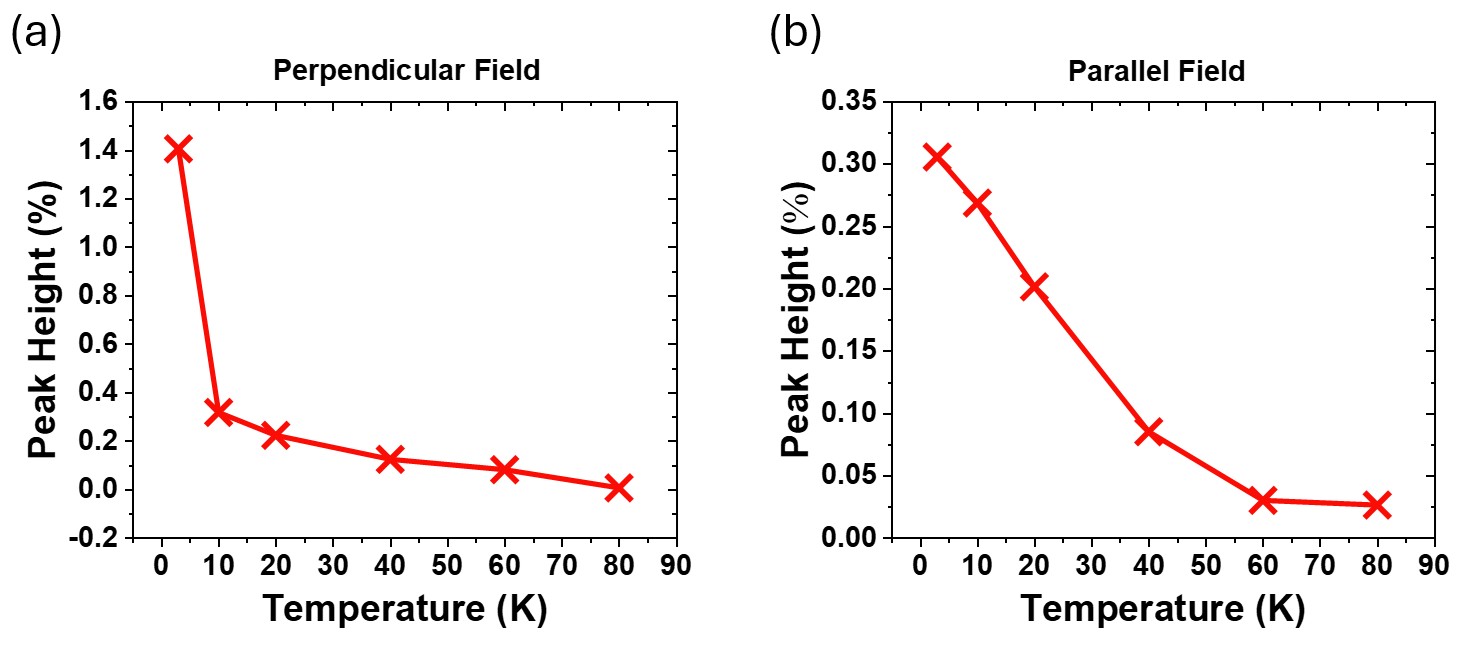}
\caption{\label{fig:nii2_s11}(a,b) Extracted heights for magnetoresistance peaks in Figs.~\ref{fig:f3}a and~\ref{fig:f3}b, respectively.}
\end{figure}

\begin{figure}[hbt!]
\centering
\includegraphics[width=\columnwidth]{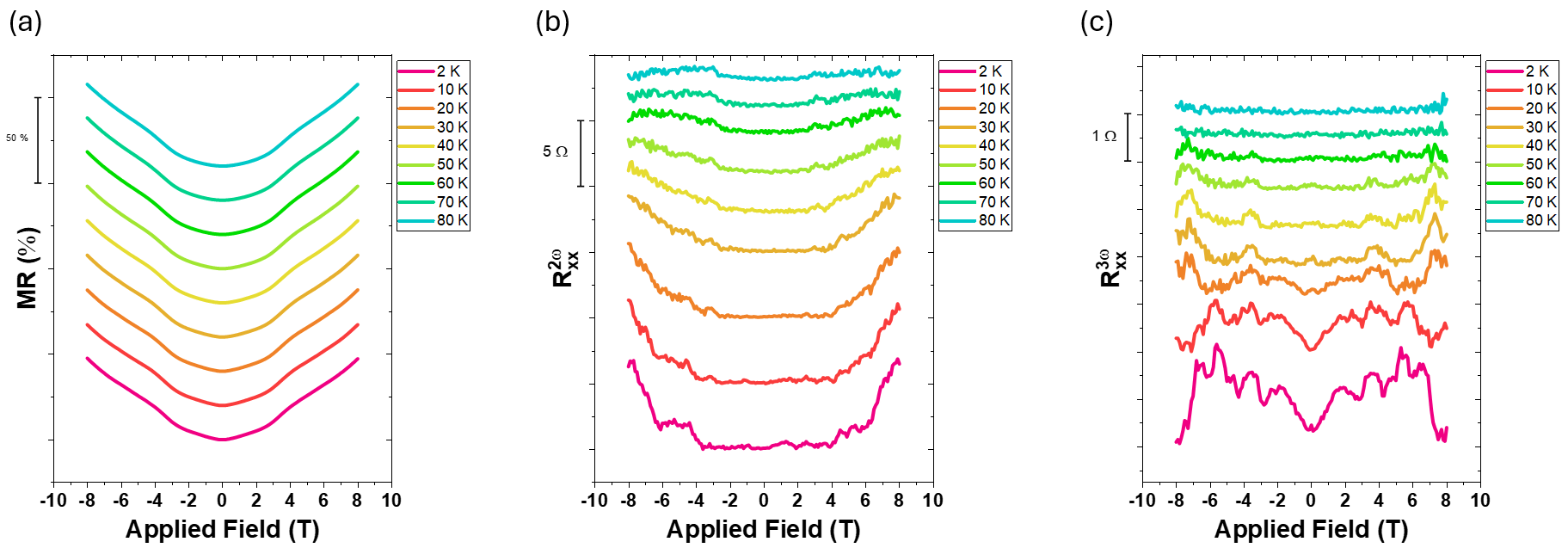}
\caption{\label{fig:nii2_s12}(a) Longitudinal magnetoresistance data for the graphene control device. (b) Second harmonic voltage of the data in (a). (c) Third harmonic voltage of the data in (a). For these measurements, the magnetic field is applied parallel to the direction of current flow.}
\end{figure}

Hall effect data for two samples. In sample 2's out-of-plane Hall response shown in Fig.~\ref{fig:nii2_s8}a, we see only a strong, linear component consistent across a wide temperature range that we would expect for graphene. No trace of \ce{NiI2}'s magnetism can be seen in the Hall response, although related features appear in the longitudinal resistance, as shown in Fig.~\ref{fig:nii2_s9}. Additionally, no trace of any proximity coupling was seen in another sample (sample 4) with an in-plane field applied at various angles, as can be seen in Fig.~\ref{fig:nii2_s8}b. For a perfectly in-plane field, we would not expect to see any signal; however, due to non-idealities such as residue from the transfer process, the applied field is not perfectly in plane, and a small response is seen.

\clearpage

Longitudinal magnetoresistance data for sample 2. As seen in Fig.~\ref{fig:nii2_s9}a, the out-of-plane response is largely parabolic, with no indication of \ce{NiI2}'s influence. This is unsurprising, considering that graphene is expected to have a strong geometric magnetoresistance. Any trace of proximity coupling is likely drowned out accordingly. In the in-plane response, however, we see the same behavior as sample 1 in the main text, albeit with a narrower peak.

\clearpage

Longitudinal magnetoresistance data for sample 3. This sample had a relatively poor SNR (likely due to multi-layer graphene), and traces of coupling to the \ce{NiI2} were not visible up to T$_{N2}$. Nevertheless, clear magnetoresistance peaks are seen, and there are detectable second and third harmonic signals. 

\clearpage

Extracted magnetoresistance peak height for data in Fig. 3. The onset of T$_{N2}$ is not as clear for (a) as it is in (b), likely due to the much wider magnetoresistance hump.

\clearpage

Longitudinal magnetoresistance data for the graphene control device. The peak features identified in the fundamental signals of the \ce{NiI2}/Gr devices are completely absent in these measurements, indicating that they are indeed from the \ce{NiI2}. We also present the second and third harmonics of the magnetoresistance data for comparison. The second harmonic remains relatively weak and does not show the pronounced peak-like structures seen in the heterostructure, while the third harmonic exhibits a stronger background response with its own field dependence. The magnetic field is applied in-plane parallel to the direction of current flow.

\clearpage

\begin{figure}[hbt!]
\centering
\includegraphics[width=\columnwidth]{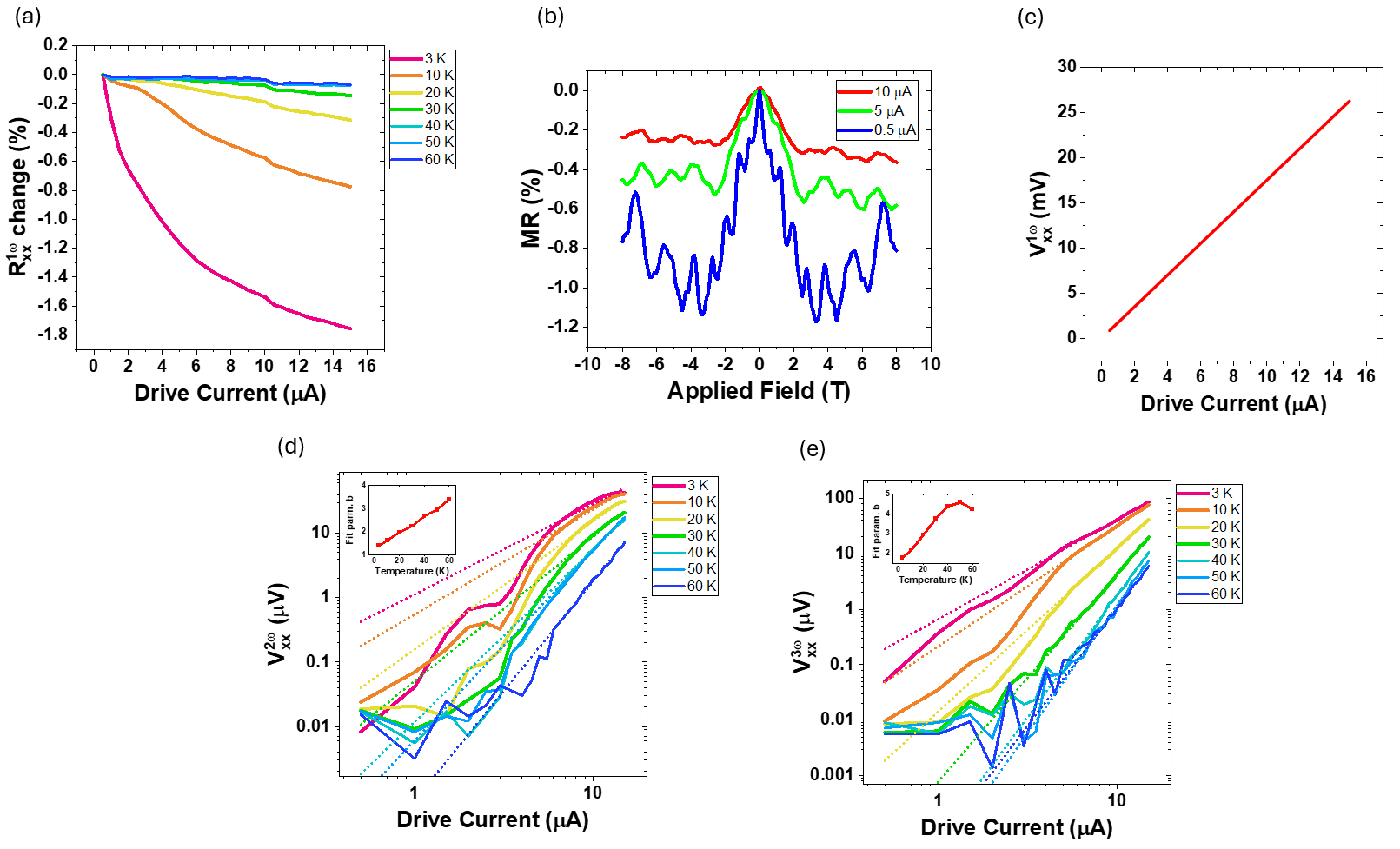}
\caption{\label{fig:nii2_s14}(a) Sample resistance vs. drive current at various temperatures. (b) Magnetoresistance sweeps at three different drive currents for a field applied parallel to the current direction. (c) First-harmonic voltage vs. drive current for sample 1 at 3 K, showing an ohmic contact. (d) Second and (e) third harmonic voltage of the data in (a) plotted on a log-log scale at various temperatures. Dotted lines represent fits. Insets: fitting exponent $b$ for each temperature series. All currents reported are RMS values.}
\end{figure}

Current-dependent measurements used to assess the role of bias in the nonlinear response are shown in Fig.~\ref{fig:nii2_s14}. Unless otherwise noted, the transport data in the main text were acquired using a drive current of 10 $\mu$A. Owing to its atomic thickness, graphene can experience substantial Joule self-heating under cryogenic conditions~\cite{gr_thermal_conductance,RN612}, with the magnitude of heating depending sensitively on the thermal properties of the full heterostructure. In Fig.~\ref{fig:nii2_s14}a, the low-temperature resistance changes appreciably with increasing drive current, indicating that the transport response is strongly bias dependent in this regime. Despite this, the first harmonic IV remains largely linear at 3 K, indicating an ohmic contact as would be expected for Gr/Au contacts.

This sensitivity is also reflected in the magnetoresistance data of Fig.~\ref{fig:nii2_s14}b. At 3 K, reducing the drive current reveals a larger magnetoresistance together with reproducible oscillatory structure that we attribute to universal conductance fluctuations (UCFs). The suppression of these features at higher drive currents shows that the measurement current can substantially modify the low-temperature transport response.

We also examine how the nonlinear signals scale with current in Fig.~\ref{fig:nii2_s14}c and d. If the response followed a simple, trivial power law over the full measured range, the data would collapse onto nearly straight lines on the log-log plots with temperature-independent exponents. Instead, both the second- and third-harmonic signals show clear curvature, and the fitted exponent $b$ varies strongly with temperature for both harmonics. We therefore take these data as evidence for non-trivial current scaling in the heterostructure. While self-heating is clearly important, the IV curves and harmonic scaling alone do not isolate a unique microscopic origin for the observed nonlinear behavior.

\clearpage

\begin{figure}[hbt!]
\centering
\includegraphics[width=\columnwidth]{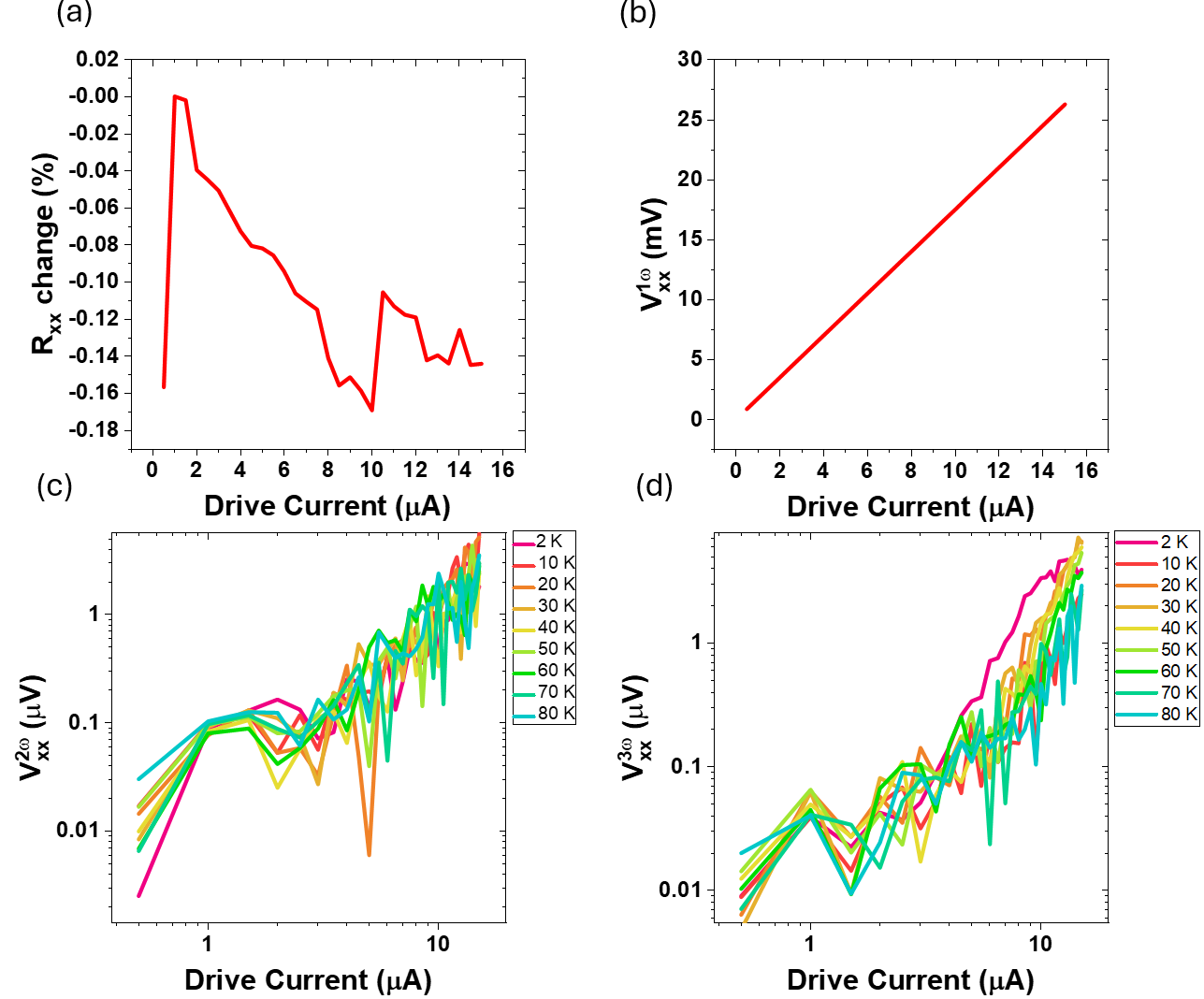}
\caption{\label{fig:nii2_s15}(a) Sample resistance vs. drive current for the graphene control device at 3 K. Higher-temperature traces are omitted for clarity because their current dependence is much smaller on this scale. (b) First-harmonic voltage vs. drive current for the graphene control device at 3 K, showing ohmic behavior. (c) Second- and (d) third-harmonic voltage vs. drive current for the control device plotted on log-log scales at various temperatures. All currents reported are RMS values.}
\end{figure}

Current-dependent measurements for the graphene control device are shown in Fig.~\ref{fig:nii2_s15}. Compared to the \ce{NiI2} heterostructure, the control device shows much weaker bias dependence in the fundamental resistance. Only the 3 K trace is shown in Fig.~\ref{fig:nii2_s15}a for clarity, since the current dependence at higher temperatures is small on this scale. The first harmonic IV is linear, as is expected for a Gr/Au contact. The second- and third-harmonic IV data in Fig.~\ref{fig:nii2_s15}b and c are also weaker and noisier than in the heterostructure and do not show the same clear temperature evolution or non-trivial scaling. This contrast suggests that the pronounced bias dependence of the \ce{NiI2} device is not a generic feature of graphene transport alone, although differences in thermal coupling between the two device stacks may also contribute.

\clearpage

\bibliographystyle{unsrtnat}
\bibliography{references}

@article{RevModPhys.90.015005,
  title = {Antiferromagnetic spintronics},
  author = {Baltz, V. and Manchon, A. and Tsoi, M. and Moriyama, T. and Ono, T. and Tserkovnyak, Y.},
  journal = {Rev. Mod. Phys.},
  volume = {90},
  issue = {1},
  pages = {015005},
  numpages = {57},
  year = {2018},
  month = {Feb},
  publisher = {American Physical Society},
  doi = {10.1103/RevModPhys.90.015005},
  url = {https://link.aps.org/doi/10.1103/RevModPhys.90.015005}
}

@article{RN596,
   author = {Rimmler, Berthold H. and Pal, Banabir and Parkin, Stuart S. P.},
   title = {Non-collinear antiferromagnetic spintronics},
   journal = {Nature Reviews Materials},
   volume = {10},
   number = {2},
   pages = {109-127},
   ISSN = {2058-8437},
   DOI = {10.1038/s41578-024-00706-w},
   url = {https://doi.org/10.1038/s41578-024-00706-w},
   year = {2025},
   type = {Journal Article}
}

@article{RN88,
   author = {Nogués, J. and Schuller, Ivan K.},
   title = {Exchange bias},
   journal = {Journal of Magnetism and Magnetic Materials},
   volume = {192},
   number = {2},
   pages = {203-232},
   DOI = {10.1016/S0304-8853(98)00266-2},
   year = {1999},
   type = {Journal Article}
}

@article{RN169,
   author = {Wadley, P. and Howells, B. and Železný, J. and Andrews, C. and Hills, V. and Campion, R. P. and Novák, V. and Olejník, K. and Maccherozzi, F. and Dhesi, S. S. and Martin, S. Y. and Wagner, T. and Wunderlich, J. and Freimuth, F. and Mokrousov, Y. and Kuneš, J. and Chauhan, J. S. and Grzybowski, M. J. and Rushforth, A. W. and Edmond, Kw and Gallagher, B. L. and Jungwirth, T.},
   title = {Electrical switching of an antiferromagnet},
   journal = {Science},
   volume = {351},
   number = {6273},
   pages = {587-590},
   DOI = {10.1126/science.aab1031},
   year = {2016},
   type = {Journal Article}
}

@article{RN597,
   author = {Bodnar, S. Yu and Šmejkal, L. and Turek, I. and Jungwirth, T. and Gomonay, O. and Sinova, J. and Sapozhnik, A. A. and Elmers, H. J. and Kläui, M. and Jourdan, M.},
   title = {Writing and reading antiferromagnetic {Mn$_2$Au} by Néel spin-orbit torques and large anisotropic magnetoresistance},
   journal = {Nature Communications},
   volume = {9},
   number = {1},
   pages = {348},
   ISSN = {2041-1723},
   DOI = {10.1038/s41467-017-02780-x},
   url = {https://doi.org/10.1038/s41467-017-02780-x},
   year = {2018},
   type = {Journal Article}
}

@article{RN598,
   author = {Olejník, Kamil and Seifert, Tom and Kašpar, Zdeněk and Novák, Vít and Wadley, Peter and Campion, Richard P. and Baumgartner, Manuel and Gambardella, Pietro and Němec, Petr and Wunderlich, Joerg and Sinova, Jairo and Kužel, Petr and Müller, Melanie and Kampfrath, Tobias and Jungwirth, Tomas},
   title = {Terahertz electrical writing speed in an antiferromagnetic memory},
   journal = {Science Advances},
   volume = {4},
   number = {3},
   pages = {eaar3566},
   DOI = {doi:10.1126/sciadv.aar3566},
   url = {https://www.science.org/doi/abs/10.1126/sciadv.aar3566},
   year = {2018},
   type = {Journal Article}
}

@article{mnbased_review,
	author={Chen, Shiwei and Meng, Dequan and Zeng, Guang and Zhang, Pan and Liang, Shiheng},
	title={Spin-Orbit-Torque Based on Mn-Based Noncollinear Antiferromagnets},
	journal={Journal of Physics: Condensed Matter},
	url={http://iopscience.iop.org/article/10.1088/1361-648X/addd54},
	year={2025},
}

@article{RN599,
   author = {Nakatsuji, Satoru and Kiyohara, Naoki and Higo, Tomoya},
   title = {Large anomalous Hall effect in a non-collinear antiferromagnet at room temperature},
   journal = {Nature},
   volume = {527},
   number = {7577},
   pages = {212-215},
   ISSN = {1476-4687},
   DOI = {10.1038/nature15723},
   url = {https://doi.org/10.1038/nature15723},
   year = {2015},
   type = {Journal Article}
}

@article{RN600,
   author = {Kimata, Motoi and Chen, Hua and Kondou, Kouta and Sugimoto, Satoshi and Muduli, Prasanta K. and Ikhlas, Muhammad and Omori, Yasutomo and Tomita, Takahiro and MacDonald, Allan H. and Nakatsuji, Satoru and Otani, Yoshichika},
   title = {Magnetic and magnetic inverse spin Hall effects in a non-collinear antiferromagnet},
   journal = {Nature},
   volume = {565},
   number = {7741},
   pages = {627-630},
   ISSN = {1476-4687},
   DOI = {10.1038/s41586-018-0853-0},
   url = {https://doi.org/10.1038/s41586-018-0853-0},
   year = {2019},
   type = {Journal Article}
}

@article{helical_thz,
  title = {Current-Driven Collective Control of Helical Spin Texture in van der Waals Antiferromagnet},
  author = {Zhang, Kai-Xuan and Cheon, Suik and Kim, Hyuncheol and Park, Pyeongjae and An, Yeochan and Son, Suhan and Cui, Jingyuan and Keum, Jihoon and Choi, Joonyoung and Jo, Younjung and Ju, Hwiin and Lee, Jong-Seok and Lee, Youjin and Avdeev, Maxim and Kleibert, Armin and Lee, Hyun-Woo and Park, Je-Geun},
  journal = {Phys. Rev. Lett.},
  volume = {134},
  issue = {17},
  pages = {176701},
  numpages = {10},
  year = {2025},
  month = {Apr},
  publisher = {American Physical Society},
  doi = {10.1103/PhysRevLett.134.176701},
  url = {https://link.aps.org/doi/10.1103/PhysRevLett.134.176701}
}

@article{first_cr2o3,
    author = {Rado, G. T. and Folen, V. J.},
    title = {Magnetoelectric Effects in Antiferromagnetics},
    journal = {Journal of Applied Physics},
    volume = {33},
    number = {3},
    pages = {1126-1132},
    year = {1962},
    month = {03},
    issn = {0021-8979},
    doi = {10.1063/1.1728630},
    url = {https://doi.org/10.1063/1.1728630},
    eprint = {https://pubs.aip.org/aip/jap/article-pdf/33/3/1126/18325864/1126\_2\_online.pdf},
}

@article{RN601,
   author = {Kosub, Tobias and Kopte, Martin and Hühne, Ruben and Appel, Patrick and Shields, Brendan and Maletinsky, Patrick and Hübner, René and Liedke, Maciej Oskar and Fassbender, Jürgen and Schmidt, Oliver G. and Makarov, Denys},
   title = {Purely antiferromagnetic magnetoelectric random access memory},
   journal = {Nature Communications},
   volume = {8},
   number = {1},
   pages = {13985},
   ISSN = {2041-1723},
   DOI = {10.1038/ncomms13985},
   url = {https://doi.org/10.1038/ncomms13985},
   year = {2017},
   type = {Journal Article}
}

@article{helicity_control_2007,
  title = {Electric Control of Spin Helicity in a Magnetic Ferroelectric},
  author = {Yamasaki, Y. and Sagayama, H. and Goto, T. and Matsuura, M. and Hirota, K. and Arima, T. and Tokura, Y.},
  journal = {Phys. Rev. Lett.},
  volume = {98},
  issue = {14},
  pages = {147204},
  numpages = {4},
  year = {2007},
  month = {Apr},
  publisher = {American Physical Society},
  doi = {10.1103/PhysRevLett.98.147204},
  url = {https://link.aps.org/doi/10.1103/PhysRevLett.98.147204}
}

@article{RN602,
   author = {Huang, Xiaoxi and Chen, Xianzhe and Li, Yuhang and Mangeri, John and Zhang, Hongrui and Ramesh, Maya and Taghinejad, Hossein and Meisenheimer, Peter and Caretta, Lucas and Susarla, Sandhya and Jain, Rakshit and Klewe, Christoph and Wang, Tianye and Chen, Rui and Hsu, Cheng-Hsiang and Harris, Isaac and Husain, Sajid and Pan, Hao and Yin, Jia and Shafer, Padraic and Qiu, Ziqiang and Rodrigues, Davi R. and Heinonen, Olle and Vasudevan, Dilip and Íñiguez, Jorge and Schlom, Darrell G. and Salahuddin, Sayeef and Martin, Lane W. and Analytis, James G. and Ralph, Daniel C. and Cheng, Ran and Yao, Zhi and Ramesh, Ramamoorthy},
   title = {Manipulating chiral spin transport with ferroelectric polarization},
   journal = {Nature Materials},
   volume = {23},
   number = {7},
   pages = {898-904},
   ISSN = {1476-4660},
   DOI = {10.1038/s41563-024-01854-8},
   url = {https://doi.org/10.1038/s41563-024-01854-8},
   year = {2024},
   type = {Journal Article}
}

@article{RN7,
   author = {Huang, Bevin and Clark, Genevieve and Navarro-Moratalla, Efrén and Klein, Dahlia R. and Cheng, Ran and Seyler, Kyle L. and Zhong, DIng and Schmidgall, Emma and McGuire, Michael A. and Cobden, David H. and Yao, Wang and Xiao, Di and Jarillo-Herrero, Pablo and Xu, Xiaodong},
   title = {Layer-dependent ferromagnetism in a van der Waals crystal down to the monolayer limit},
   journal = {Nature},
   volume = {546},
   number = {7657},
   pages = {270-273},
   DOI = {10.1038/nature22391},
   url = {https://www.nature.com/articles/nature22391.pdf},
   year = {2017},
   type = {Journal Article}
}

@article{RN116,
   author = {Li, Wei and Zeng, Yi and Zhao, Zijing and Zhang, Biao and Xu, Junjie and Huang, Xiaoxiao and Hou, Yanglong},
   title = {2D Magnetic Heterostructures and Their Interface Modulated Magnetism},
   journal = {ACS Applied Materials and Interfaces},
   volume = {13},
   number = {43},
   pages = {50591-50601},
   DOI = {10.1021/acsami.1c11132},
   year = {2021},
   type = {Journal Article}
}

@article{RN526,
   author = {Yang, H. and Valenzuela, S. O. and Chshiev, M. and Couet, S. and Dieny, B. and Dlubak, B. and Fert, A. and Garello, K. and Jamet, M. and Jeong, D. E. and Lee, K. and Lee, T. and Martin, M. B. and Kar, G. S. and Seneor, P. and Shin, H. J. and Roche, S.},
   title = {Two-dimensional materials prospects for non-volatile spintronic memories},
   journal = {Nature},
   volume = {606},
   number = {7915},
   pages = {663-673},
   ISSN = {1476-4687 (Electronic) 0028-0836 (Linking)},
   DOI = {10.1038/s41586-022-04768-0},
   url = {https://www.ncbi.nlm.nih.gov/pubmed/35732761},
   year = {2022},
   type = {Journal Article}
}

@article{RN603,
   author = {Rahman, Sharidya and Torres, Juan F. and Khan, Ahmed Raza and Lu, Yuerui},
   title = {Recent Developments in van der Waals Antiferromagnetic 2D Materials: Synthesis, Characterization, and Device Implementation},
   journal = {ACS Nano},
   volume = {15},
   number = {11},
   pages = {17175-17213},
   note = {doi: 10.1021/acsnano.1c06864},
   ISSN = {1936-0851},
   DOI = {10.1021/acsnano.1c06864},
   url = {https://doi.org/10.1021/acsnano.1c06864},
   year = {2021},
   type = {Journal Article}
}

@article{RN604,
   author = {Sun, Yue and Meng, Fanhao and Lee, Changmin and Soll, Aljoscha and Zhang, Hongrui and Ramesh, Ramamoorthy and Yao, Jie and Sofer, Zdeněk and Orenstein, Joseph},
   title = {Dipolar spin wave packet transport in a van der Waals antiferromagnet},
   journal = {Nature Physics},
   volume = {20},
   number = {5},
   pages = {794-800},
   ISSN = {1745-2481},
   DOI = {10.1038/s41567-024-02387-2},
   url = {https://doi.org/10.1038/s41567-024-02387-2},
   year = {2024},
   type = {Journal Article}
}

@article{nii2_foundational,
title = {Magnetic and structural investigations on {NiI$_2$} and {CoI$_2$}},
journal = {Physica B+C},
volume = {111},
number = {2},
pages = {231-248},
year = {1981},
issn = {0378-4363},
doi = {https://doi.org/10.1016/0378-4363(81)90100-5},
url = {https://www.sciencedirect.com/science/article/pii/0378436381901005},
author = {S.R. Kuindersma and J.P. Sanchez and C. Haas}
}

@article{nii2_transport,
author = {Lebedev, Dmitry and Gish, Jonathan Tyler and Garvey, Ethan Skyler and Stanev, Teodor Kosev and Choi, Junhwan and Georgopoulos, Leonidas and Song, Thomas Wei and Park, Hong Youl and Watanabe, Kenji and Taniguchi, Takashi and Stern, Nathaniel Patrick and Sangwan, Vinod Kumar and Hersam, Mark Christopher},
title = {Electrical Interrogation of Thickness-Dependent Multiferroic Phase Transitions in the 2D Antiferromagnetic Semiconductor {NiI$_2$}},
journal = {Advanced Functional Materials},
volume = {33},
number = {12},
pages = {2212568},
doi = {https://doi.org/10.1002/adfm.202212568},
url = {https://advanced.onlinelibrary.wiley.com/doi/abs/10.1002/adfm.202212568},
eprint = {https://advanced.onlinelibrary.wiley.com/doi/pdf/10.1002/adfm.202212568},
year = {2023}
}

@article{RN606,
   author = {Wu, Yangliu and Zeng, Zhaozhuo and Lu, Haipeng and Han, Xiaocang and Yang, Chendi and Liu, Nanshu and Zhao, Xiaoxu and Qiao, Liang and Ji, Wei and Che, Renchao and Deng, Longjiang and Yan, Peng and Peng, Bo},
   title = {Coexistence of ferroelectricity and antiferroelectricity in 2D van der Waals multiferroic},
   journal = {Nature Communications},
   volume = {15},
   number = {1},
   pages = {8616},
   ISSN = {2041-1723},
   DOI = {10.1038/s41467-024-53019-5},
   url = {https://doi.org/10.1038/s41467-024-53019-5},
   year = {2024},
   type = {Journal Article}
}

@article{RN605,
   author = {Liu, Haining and Wang, Xinsheng and Wu, Juanxia and Chen, Yuansha and Wan, Jing and Wen, Rui and Yang, Jinbo and Liu, Ying and Song, Zhigang and Xie, Liming},
   title = {Vapor Deposition of Magnetic Van der Waals {NiI$_2$} Crystals},
   journal = {ACS Nano},
   volume = {14},
   number = {8},
   pages = {10544-10551},
   note = {doi: 10.1021/acsnano.0c04499},
   ISSN = {1936-0851},
   DOI = {10.1021/acsnano.0c04499},
   url = {https://doi.org/10.1021/acsnano.0c04499},
   year = {2020},
   type = {Journal Article}
}

@article{RN133,
   author = {Huang, S. Y. and Fan, X. and Qu, D. and Chen, Y. P. and Wang, W. G. and Wu, J. and Chen, T. Y. and Xiao, J. Q. and Chien, C. L.},
   title = {Transport magnetic proximity effects in platinum},
   journal = {Physical Review Letters},
   volume = {109},
   number = {10},
   pages = {107204},
   DOI = {10.1103/PhysRevLett.109.107204},
   url = {https://journals.aps.org/prl/pdf/10.1103/PhysRevLett.109.107204},
   year = {2012},
   type = {Journal Article}
}

@article{RN543,
   author = {Lohmann, M. and Su, T. and Niu, B. and Hou, Y. and Alghamdi, M. and Aldosary, M. and Xing, W. and Zhong, J. and Jia, S. and Han, W. and Wu, R. and Cui, Y. T. and Shi, J.},
   title = {Probing Magnetism in Insulating {Cr$_{2}$Ge$_{2}$Te$_{6}$} by Induced Anomalous Hall Effect in Pt},
   journal = {Nano Lett},
   volume = {19},
   number = {4},
   pages = {2397-2403},
   ISSN = {1530-6992 (Electronic) 1530-6984 (Linking)},
   DOI = {10.1021/acs.nanolett.8b05121},
   url = {https://www.ncbi.nlm.nih.gov/pubmed/30823703},
   year = {2019},
   type = {Journal Article}
}

@article{RN568,
   author = {Tang, C. and Zhang, Z. and Lai, S. and Tan, Q. and Gao, W. B.},
   title = {Magnetic Proximity Effect in {Graphene/CrBr$_3$} van der Waals Heterostructures},
   journal = {Adv Mater},
   volume = {32},
   number = {16},
   pages = {e1908498},
   ISSN = {1521-4095 (Electronic)
0935-9648 (Linking)},
   DOI = {10.1002/adma.201908498},
   url = {https://www.ncbi.nlm.nih.gov/pubmed/32130750},
   year = {2020},
   type = {Journal Article}
}

@article{RN201,
   author = {Purdie, D. G. and Pugno, N. M. and Taniguchi, T. and Watanabe, K. and Ferrari, A. C. and Lombardo, A.},
   title = {Cleaning interfaces in layered materials heterostructures},
   journal = {Nat Commun},
   volume = {9},
   number = {1},
   pages = {5387},
   note = {Purdie, D G
Pugno, N M
Taniguchi, T
Watanabe, K
Ferrari, A C
Lombardo, A
eng
Research Support, Non-U.S. Gov't
England
2018/12/21
Nat Commun. 2018 Dec 19;9(1):5387. doi: 10.1038/s41467-018-07558-3.},
   ISSN = {2041-1723 (Electronic)
2041-1723 (Linking)},
   DOI = {10.1038/s41467-018-07558-3},
   url = {https://www.ncbi.nlm.nih.gov/pubmed/30568160},
   year = {2018},
   type = {Journal Article}
}

@article{RN607,
   author = {Wang, Xiaolei and Shang, Zixuan and Zhang, Chen and Kang, Jiaqian and Liu, Tao and Wang, Xueyun and Chen, Siliang and Liu, Haoliang and Tang, Wei and Zeng, Yu-Jia and Guo, Jianfeng and Cheng, Zhihai and Liu, Lei and Pan, Dong and Tong, Shucheng and Wu, Bo and Xie, Yiyang and Wang, Guangcheng and Deng, Jinxiang and Zhai, Tianrui and Deng, Hui-Xiong and Hong, Jiawang and Zhao, Jianhua},
   title = {Electrical and magnetic anisotropies in van der Waals multiferroic CuCrP2S6},
   journal = {Nature Communications},
   volume = {14},
   number = {1},
   pages = {840},
   ISSN = {2041-1723},
   DOI = {10.1038/s41467-023-36512-1},
   url = {https://doi.org/10.1038/s41467-023-36512-1},
   year = {2023},
   type = {Journal Article}
}

@article{graphene_hg_review,
title = {Engineering the harmonic generation in graphene},
journal = {Materials Today Physics},
volume = {23},
pages = {100649},
year = {2022},
issn = {2542-5293},
doi = {https://doi.org/10.1016/j.mtphys.2022.100649},
url = {https://www.sciencedirect.com/science/article/pii/S2542529322000475},
author = {R. Zhou and T. Guo and L. Huang and K. Ullah}
}

@article{RN608,
   author = {Soavi, Giancarlo and Wang, Gang and Rostami, Habib and Purdie, David G. and De Fazio, Domenico and Ma, Teng and Luo, Birong and Wang, Junjia and Ott, Anna K. and Yoon, Duhee and Bourelle, Sean A. and Muench, Jakob E. and Goykhman, Ilya and Dal Conte, Stefano and Celebrano, Michele and Tomadin, Andrea and Polini, Marco and Cerullo, Giulio and Ferrari, Andrea C.},
   title = {Broadband, electrically tunable third-harmonic generation in graphene},
   journal = {Nature Nanotechnology},
   volume = {13},
   number = {7},
   pages = {583-588},
   ISSN = {1748-3395},
   DOI = {10.1038/s41565-018-0145-8},
   url = {https://doi.org/10.1038/s41565-018-0145-8},
   year = {2018},
   type = {Journal Article}
}

@article{RN609,
   author = {Zhang, Naiyuan James and Lin, Jiang-Xiazi and Chichinadze, Dmitry V. and Wang, Yibang and Watanabe, Kenji and Taniguchi, Takashi and Fu, Liang and Li, J. I. A.},
   title = {Angle-resolved transport non-reciprocity and spontaneous symmetry breaking in twisted trilayer graphene},
   journal = {Nature Materials},
   volume = {23},
   number = {3},
   pages = {356-362},
   ISSN = {1476-4660},
   DOI = {10.1038/s41563-024-01809-z},
   url = {https://doi.org/10.1038/s41563-024-01809-z},
   year = {2024},
   type = {Journal Article}
}

@article{nonlinear_skyrmion,
  title = {Strongly pinned skyrmionic bubbles and higher-order nonlinear Hall resistances at the interface of Pt/FeSi bilayer},
  author = {Hori, T. and Kanazawa, N. and Matsuura, K. and Ishizuka, H. and Fujiwara, K. and Tsukazaki, A. and Ichikawa, M. and Kawasaki, M. and Kagawa, F. and Hirayama, M. and Tokura, Y.},
  journal = {Phys. Rev. Mater.},
  volume = {8},
  issue = {4},
  pages = {044407},
  numpages = {8},
  year = {2024},
  month = {Apr},
  publisher = {American Physical Society},
  doi = {10.1103/PhysRevMaterials.8.044407},
  url = {https://link.aps.org/doi/10.1103/PhysRevMaterials.8.044407}
}

@article{RN487,
   author = {Song, Q. and Occhialini, C. A. and Ergecen, E. and Ilyas, B. and Amoroso, D. and Barone, P. and Kapeghian, J. and Watanabe, K. and Taniguchi, T. and Botana, A. S. and Picozzi, S. and Gedik, N. and Comin, R.},
   title = {Evidence for a single-layer van der Waals multiferroic},
   journal = {Nature},
   volume = {602},
   number = {7898},
   pages = {601-605},
   note = {Song, Qian
Occhialini, Connor A
Ergecen, Emre
Ilyas, Batyr
Amoroso, Danila
Barone, Paolo
Kapeghian, Jesse
Watanabe, Kenji
Taniguchi, Takashi
Botana, Antia S
Picozzi, Silvia
Gedik, Nuh
Comin, Riccardo
eng
Research Support, Non-U.S. Gov't
Research Support, U.S. Gov't, Non-P.H.S.
England
2022/02/25
Nature. 2022 Feb;602(7898):601-605. doi: 10.1038/s41586-021-04337-x. Epub 2022 Feb 23.},
   ISSN = {1476-4687 (Electronic)
0028-0836 (Linking)},
   DOI = {10.1038/s41586-021-04337-x},
   url = {https://www.ncbi.nlm.nih.gov/pubmed/35197619},
   year = {2022},
   type = {Journal Article}
}

@article{graphene_3w_thermal,
    author = {Chen, Z. and Jang, W. and Bao, W. and Lau, C. N. and Dames, C.},
    title = {Thermal contact resistance between graphene and silicon dioxide},
    journal = {Applied Physics Letters},
    volume = {95},
    number = {16},
    pages = {161910},
    year = {2009},
    month = {10},
    issn = {0003-6951},
    doi = {10.1063/1.3245315},
    url = {https://doi.org/10.1063/1.3245315},
    eprint = {https://pubs.aip.org/aip/apl/article-pdf/doi/10.1063/1.3245315/14423098/161910\_1\_online.pdf},
}

@article{fm_hm_3w,
  title = {Third harmonic resistance in ferromagnet/heavy-metal bilayers},
  author = {Ma, Yuteng and Xie, Hang and Si, Yuxin and Rong, Bin and Wang, Jiaqi and Zheng, Hongsheng and Liu, Yanghui and Wu, Yihong and Yang, Yumeng},
  journal = {Phys. Rev. B},
  volume = {111},
  issue = {9},
  pages = {094409},
  numpages = {11},
  year = {2025},
  month = {Mar},
  publisher = {American Physical Society},
  doi = {10.1103/PhysRevB.111.094409},
  url = {https://link.aps.org/doi/10.1103/PhysRevB.111.094409}
}

@article{3w_method,
    author = {Cahill, David G.},
    title = {Thermal conductivity measurement from 30 to 750 K: the 3$\omega$ method},
    journal = {Review of Scientific Instruments},
    volume = {61},
    number = {2},
    pages = {802-808},
    year = {1990},
    month = {02},
    issn = {0034-6748},
    doi = {10.1063/1.1141498},
    url = {https://doi.org/10.1063/1.1141498},
    eprint = {https://pubs.aip.org/aip/rsi/article-pdf/61/2/802/19111597/802\_1\_online.pdf},
}

@article{RN352,
   author = {Blake, P. and Hill, E. W. and Castro Neto, A. H. and Novoselov, K. S. and Jiang, D. and Yang, R. and Booth, T. J. and Geim, A. K.},
    title = {Making graphene visible},
    journal = {Applied Physics Letters},
    volume = {91},
    number = {6},
    pages = {063124},
    year = {2007},
    month = {08},
    issn = {0003-6951},
    doi = {10.1063/1.2768624},
    url = {https://doi.org/10.1063/1.2768624},
    eprint = {https://pubs.aip.org/aip/apl/article-pdf/doi/10.1063/1.2768624/13197090/063124\_1\_online.pdf},
}

@article{RN610,
   author = {Kim, Jong Hyuk and Shin, Hyun Jun and Kim, Mi Kyung and Hong, Jae Min and Jeong, Ki Won and Kim, Jin Seok and Moon, Kyungsun and Lee, Nara and Choi, Young Jai},
   title = {Sign-tunable anisotropic magnetoresistance and electrically detectable dual magnetic phases in a helical antiferromagnet},
   journal = {NPG Asia Materials},
   volume = {14},
   number = {1},
   pages = {67},
   ISSN = {1884-4057},
   DOI = {10.1038/s41427-022-00415-2},
   url = {https://doi.org/10.1038/s41427-022-00415-2},
   year = {2022},
   type = {Journal Article}
}

@article{helical_afm_theory,
  title = {Magnetic structure and magnetization of helical antiferromagnets in high magnetic fields perpendicular to the helix axis at zero temperature},
  author = {Johnston, David C.},
  journal = {Phys. Rev. B},
  volume = {96},
  issue = {10},
  pages = {104405},
  numpages = {22},
  year = {2017},
  month = {Sep},
  publisher = {American Physical Society},
  doi = {10.1103/PhysRevB.96.104405},
  url = {https://link.aps.org/doi/10.1103/PhysRevB.96.104405}
}

@article{cgt_graphene_prox,
doi = {10.1088/2053-1583/ab5915},
url = {https://dx.doi.org/10.1088/2053-1583/ab5915},
year = {2019},
month = {dec},
publisher = {IOP Publishing},
volume = {7},
number = {1},
pages = {015026},
author = {Karpiak, Bogdan and Cummings, Aron W and Zollner, Klaus and Vila, Marc and Khokhriakov, Dmitrii and Hoque, Anamul Md and Dankert, André and Svedlindh, Peter and Fabian, Jaroslav and Roche, Stephan and Dash, Saroj P},
title = {Magnetic proximity in a van der Waals heterostructure of magnetic insulator and graphene},
journal = {2D Materials}
}

@article{helimagnet_phase_transition,
  title = {Magnetic phase transition in single crystals of the chiral helimagnet Cr${}_{1/3}$NbS${}_{2}$},
  author = {Ghimire, N. J. and McGuire, M. A. and Parker, D. S. and Sipos, B. and Tang, S. and Yan, J.-Q. and Sales, B. C. and Mandrus, D.},
  journal = {Phys. Rev. B},
  volume = {87},
  issue = {10},
  pages = {104403},
  numpages = {9},
  year = {2013},
  month = {Mar},
  publisher = {American Physical Society},
  doi = {10.1103/PhysRevB.87.104403},
  url = {https://link.aps.org/doi/10.1103/PhysRevB.87.104403}
}

@article{shear_mediated,
author = {Tseng, Yi and Occhialini, Connor A. and Song, Qian and Barone, Paolo and Patel, Sahaj and Shankar, Meghna and Acevedo-Esteves, Raul and Li, Jiarui and Nelson, Christie and Picozzi, Silvia and Sutarto, Ronny and Comin, Riccardo},
title = {Shear-Mediated Stabilization of Spin Spiral Order in Multiferroic NiI2},
journal = {Advanced Materials},
volume = {37},
number = {9},
pages = {2417434},
doi = {https://doi.org/10.1002/adma.202417434},
url = {https://advanced.onlinelibrary.wiley.com/doi/abs/10.1002/adma.202417434},
eprint = {https://advanced.onlinelibrary.wiley.com/doi/pdf/10.1002/adma.202417434},
year = {2025}
}

@article{RN611,
   author = {Liu, Qiye and Su, Wenjie and Gu, Yue and Zhang, Xi and Xia, Xiuquan and Wang, Le and Xiao, Ke and Zhang, Naipeng and Cui, Xiaodong and Huang, Mingyuan and Wei, Chengrong and Zou, Xiaolong and Xi, Bin and Mei, Jia-Wei and Dai, Jun-Feng},
   title = {Surprising pressure-induced magnetic transformations from helimagnetic order to antiferromagnetic state in NiI2},
   journal = {Nature Communications},
   volume = {16},
   number = {1},
   pages = {4221},
   ISSN = {2041-1723},
   DOI = {10.1038/s41467-025-59561-0},
   url = {https://doi.org/10.1038/s41467-025-59561-0},
   year = {2025},
   type = {Journal Article}
}

@article{gr_thermal_conductance,
  title = {Measurement of the Electronic Thermal Conductance Channels and Heat Capacity of Graphene at Low Temperature},
  author = {Fong, Kin Chung and Wollman, Emma E. and Ravi, Harish and Chen, Wei and Clerk, Aashish A. and Shaw, M. D. and Leduc, H. G. and Schwab, K. C.},
  journal = {Phys. Rev. X},
  volume = {3},
  issue = {4},
  pages = {041008},
  numpages = {7},
  year = {2013},
  month = {Oct},
  publisher = {American Physical Society},
  doi = {10.1103/PhysRevX.3.041008},
  url = {https://link.aps.org/doi/10.1103/PhysRevX.3.041008}
}

@article{RN612,
   author = {Koh, Yee Kan and Bae, Myung-Ho and Cahill, David G. and Pop, Eric},
   title = {Heat Conduction across Monolayer and Few-Layer Graphenes},
   journal = {Nano Letters},
   volume = {10},
   number = {11},
   pages = {4363-4368},
   note = {doi: 10.1021/nl101790k},
   ISSN = {1530-6984},
   DOI = {10.1021/nl101790k},
   url = {https://doi.org/10.1021/nl101790k},
   year = {2010},
   type = {Journal Article}
}

@article{dames2013measuring,
  title={Measuring the thermal conductivity of thin films: 3 omega and related electrothermal methods},
  author={Dames, Chris},
  journal={Annual Review of Heat Transfer},
  volume={16},
  year={2013},
  publisher={Begel House Inc.}
}

\end{document}